\documentclass{emulateapj}

\usepackage{amsmath}
\usepackage{epsfig}
\usepackage{amssymb}
\usepackage{graphicx}
\usepackage{color}
\usepackage{natbib}

\def\gtaprx {\lower .1ex\hbox{\rlap{\raise .6ex\hbox{\hskip .3ex
	{\ifmmode{\scriptscriptstyle >}\else
		{$\scriptscriptstyle >$}\fi}}}
	\kern -.4ex{\ifmmode{\scriptscriptstyle \sim}\else
		{$\scriptscriptstyle\sim$}\fi}}}
\def\ltaprx {\lower .1ex\hbox{\rlap{\raise .6ex\hbox{\hskip .3ex
	{\ifmmode{\scriptscriptstyle <}\else
		{$\scriptscriptstyle <$}\fi}}}
	\kern -.4ex{\ifmmode{\scriptscriptstyle \sim}\else
		{$\scriptscriptstyle\sim$}\fi}}}

\newcommand{\cutt}[1]{\textcolor{blue}{}}

\newcommand{\Ms}{{\ensuremath{M_{\odot} }}}
\newcommand{\Zs}{\ensuremath{Z_\odot}}

\newcommand{\Ni}{{\ensuremath{^{56}\mathrm{Ni}}}}

\begin{document}

\title{Pair-Instability Supernovae in the Local Universe}

\author{Daniel J. Whalen\altaffilmark{1}, Joseph Smidt\altaffilmark{2}, Alexander 
Heger\altaffilmark{3,4}, Raphael Hirschi\altaffilmark{5,6}, Norhasliza Yusof\altaffilmark{7}, 
Wesley Even\altaffilmark{8}, Chris L. Fryer\altaffilmark{8}, Massimo Stiavelli\altaffilmark{9}, 
Ke-Jung Chen\altaffilmark{10}, and Candace C. Joggerst\altaffilmark{11}}

\altaffiltext{1}{Zentrum f\"{u}r Astronomie, Institut f\"{u}r Theoretische Astrophysik, 
Universit\"{a}t Heidelberg, Albert-Ueberle-Str. 2, 69120 Heidelberg, Germany}

\altaffiltext{2}{T-2, Los Alamos National Laboratory, Los Alamos, NM 87545, USA}

\altaffiltext{3}{Monash Centre for Astrophysics, Monash University, Victoria, 3800, 
Australia}

\altaffiltext{4}{School of Physics and Astronomy, University of Minnesota, Minneapolis, 
MN 55455, USA}

\altaffiltext{5}{Astrophysics Group, University of Keele, Lennard-Jones Labs, Keele 
ST5 5BG, UK}

\altaffiltext{6}{Institute for the Physics and Mathematics of the Universe (WPI), 
University of Tokyo, 5-1-5 Kashiwanoha, Kashiwa 277-8583, Japan}

\altaffiltext{7}{Department of Physics, University of Malaya, 50603 Kuala Lumpur, 
Malaysia}

\altaffiltext{8}{CCS-2, Los Alamos National Laboratory, Los Alamos, NM 87545, USA}

\altaffiltext{9}{Space Telescope Science Institute, 3700 San Martin Drive, Baltimore, 
MD 21218, USA}

\altaffiltext{10}{Department of Astronomy and Astrophysics, UCSC, Santa Cruz, CA  
95064, USA}

\altaffiltext{11}{XTD-3, Los Alamos National Laboratory, Los Alamos, NM 87545, 
USA}

\begin{abstract}

The discovery of 150 - 300 \Ms\ stars in the Local Group and pair-instability supernova 
candidates at low redshifts has excited interest in this exotic explosion mechanism.  
Realistic light curves for pair-instability supernovae at near-solar metallicities are key to 
identifying and properly interpreting these events as more are found. We have modeled 
pair-instability supernovae of 150 - 500 \Ms\ $Z \sim$ 0.1 - 0.4 \Zs\ stars.  These stars 
lose up to 80\% of their mass to strong line-driven winds and explode as bare He cores.  
We find that their light curves and spectra are quite different from those of Population III 
pair-instability explosions, which therefore cannot be used as templates for low-redshift 
events.  Although non-zero metallicity pair-instability supernovae are generally dimmer 
than their Population III counterparts, in some cases they will be bright enough to be 
detected at the earliest epochs at which they can occur, the formation of the first 
galaxies at $z \sim$ 10 - 15. Others can masquerade as dim, short duration supernovae 
that are only visible in the local universe and that under the right conditions could be 
hidden in a wide variety of supernova classes. We also report for the first time that some 
pair-instability explosions can create black holes with masses of $\sim$ 100 \Ms.

\end{abstract}

\keywords{early universe -- galaxies: high-redshift -- stars: early-type -- supernovae: general -- 
radiative transfer -- hydrodynamics -- cosmology:theory -- stars: Population II -- supernovae: 
individual (SN 2007bi)}

\section{Introduction}

The recent discovery of stars with masses above 150 \Ms\ in the star cluster R136 \citep{
R136} and the detection of pair-instability supernova (PI SN) candidates SN 2007bi at $z 
=$ 0.123 \citep{gy09,yn10} and SN 2213 - 1745 at $z =$ 2.05 \citep{cooke12} have excited 
interest in this exotic explosion mechanism and challenged current theories of galactic star 
formation.  In particular, it is not understood how so massive a progenitor can form in 
metallicities like those of the Small Magellanic Cloud, $\sim$ 0.1 \Zs.  Models of Population 
III (Pop III) stellar evolution predict that metal-free stars must have initial masses of 140 - 
260 \Ms\ to die as PI SNe \citep{hw02} \citep[although][have now extended the lower mass 
limit down to 85 \Ms\ if the star is rotating]{cw12}.  Stars forming in $Z \sim$ 0.1 \Zs\ gas 
may have to be much more massive at birth to die as PI SNe because they can lose most 
of their mass over their lifetimes to line-driven winds \citep{s02,Vink01}.  The conundrum 
lies in the fact that radiative cooling in gas and dust at such metallicities would almost 
certainly cause it to fragment on mass scales that are an order of magnitude below those 
required to produce PI SNe.  

Understanding the observational signatures of PI SNe at near-solar metallicities is key to 
identifying and properly interpreting these events as more of them are found, since current 
and future SN factories like the Palomar Transient Factory \citep[PTF;][]{ptf}, the Panoramic 
Survey Telescope \& Rapid Response System (Pan-STARRS) \citep{panstarrs}, and the 
Large Synoptic Survey Telescope \citep[LSST;][]{lsst} may harvest large numbers of them.  
Until now, numerical studies have focused on Pop III PI SNe as probes of the properties of 
the first stars \citep{herz90,byh03,sc05,ky05,wet08a,fwf10,kasen11,ds11a,vas12,pan12a,
pan12b,mw12,wet12b,wet12a,ds13,ds14} \citep[see also][]{wet12e,wet12d,wet12c,jet13a,
wet13b,wet13a,wet13d,wet13c,kz14a,kz14b}.  The goal of these studies was also to better 
understand primeval galaxies and the origin of supermassive black holes \citep[e.g.,][]{wf12,
jlj12a,jet13,jet14} \citep[see][for recent reviews]{dw12,glov12}.

But to understand PI SNe in the local universe, one must begin with non-zero 
metallicity progenitors for two reasons. First, metals substantially alter the internal structure 
of massive stars and can lead to heavy mass loss.  Mass loss in turn can shorten the 
duration of SN light curves (LCs) by reducing the mass of \Ni\ formed in the explosion and 
radiation diffusion timescales in the ejecta \citep{kasen11}:
\begin{equation}
t_d \sim \kappa^{\frac{1}{2}} {M_{\mathrm{ej}}}^{\frac{3}{4}} E^{-\frac{3}{4}}. \label{eq:rdt}
\end{equation}
On the other hand, the presence of a dense circumstellar envelope can add to the 
diffusion timescales of photon diffusion from the flow by contributing additional mass in
equation (\ref{eq:rdt}).  Second, Pop III stars explode in low-density relic H II regions 
\citep{wan04,ket04,oet05,abs06,awb07,wa08a,wn08b,wn08a,wet08b,wet10} while PI 
SNe in the local universe may occur in dense winds and bubbles blown by the star.  
Such envelopes can have large effects on the light curves of SNe, either quenching them 
or brightening them.  For example, outbursts prior to the death of the star can eject 
massive shells with which the SN later collides, producing an event that is far more 
luminous than the explosion itself. The effects of higher metallicity on radiation flow in the 
ejecta and surrounding interstellar medium (ISM) are also unknown.  They could be 
important, since variations in opacity of just a factor of 3 or 4 can lead to fluctuations in 
optical depth and luminosity by factors of 50 or more.

We have now modeled the PI SN explosions of very massive $Z \sim $ 0.1 - 0.4 \Zs\ stars and 
calculated their light curves and spectra.  In Section 2 we discuss our stellar evolution and SN 
models, how the SNe were subsequently evolved, and how spectra were obtained from their 
blast profiles.  We study explosions in dense envelopes in Section 3 and explosions in diffuse 
environments in Section 4.  Fallback and black hole formation are examined in Section 5, and 
we conclude in Section 6.

\section{Numerical Algorithm}

We compute light curves and spectra in four stages.  First, 150 - 500 \Ms\ stars are evolved
from the beginning of the main sequence to the end of He burning and onset of the PI in the 
GENEVA code.  We then explode the star in the Kepler code, solving for all nuclear burning 
and evolving the shock to just below the surface of the star.  Shock breakout and expansion 
into the wind are then modeled with the RAGE radiation hydrodynamics code.  Finally, we 
post process blast profiles from RAGE with the SPECTRUM code to compute spectra and 
construct light curves.  

\subsection{GENEVA Stellar Evolution Model}

We evolved 150, 200, and 500 \Ms\ stars in the latest version of the GENEVA code \citep{
hmm04,e08}.  Our simulations include mass loss and stellar rotation, which are crucial to 
the evolution of very massive stars.  We employ the theoretical prescription for mass loss 
from O stars of \citet{Vink01} \citep[which have been empirically corroborated by][]{met07} 
up to the onset of the Wolf-Rayet (WR) phase.  The WR phase is taken to begin when the 
H mass fraction $\chi_H$ at the surface of the star falls below 5\% when $T_{eff} >$ 10$^
4$ K, at which point we transition to empirical mass loss rates \citep{nl00}.  We describe 
our prescription for mass loss in greater detail in \citet{yus13}.  Three of our four models 
(h150, h200, h500s4) have an initial velocity,  $v_{\mathrm{init}} =$ 0.4 $v_{\mathrm{crit}}$, 
where $v_{\mathrm{crit}}$ is the critical (breakup) velocity. This corresponds to an average 
velocity on the main-sequence of 318.9, 333.4 and 116.8 km/s for the 150, 200 and 500 
\Ms\ stars, respectively. The fourth model (h500s0) is non-rotating.  The main effects of 
rotation are included in our calculations:  centrifugal support, mass-loss enhancement and 
especially mixing in radiative zones \citep{m09}.  The 150 and 200 \Ms\ stars have $Z \sim
$ 0.14 \Zs\ corresponding to those of the Small Magellanic Cloud.  The two 500 \Ms\ stars 
have metallicities $Z \sim$ 0.43 \Zs\ corresponding to those of the Large Magellanic Cloud.
By the end of their evolution in GENEVA all four stars have shed their hydrogen envelopes
and are essentially bare He cores.
  
\subsection{KEPLER Burn Models}

The star is evolved from the end of core He burning through explosive O and Si burning and 
then almost to shock breakout in the one-dimensional (1D) Kepler code \citep{Weaver1978,
Woosley2002}.  The explosion is the natural endpoint of stellar evolution in Kepler and is not 
artificially triggered.  Its energy is set by how much O and Si burns, which takes $\sim$ 20 s.  
We calculate energy production with a 19-isotope network up to the point of oxygen depletion 
in the core and with a 128-isotope quasi-equilibrium network thereafter. The number of mass 
zones in the models varied from 1000 - 1200 and was always sufficient to resolve the SN and 
surrounding star.  We summarize the properties of our PI SNe in Table~\ref{tab:t1}.  Their 
explosion energies vary from 3.7 to 65 B (1 B = 10$^{51}$ erg).  The s0 and s4 designations 
in Table \ref{tab:t1} refer to non-rotating and rotating stars, respectively.

\subsection{RAGE Simulations}

\begin{deluxetable}{lccccc}  
\tabletypesize{\scriptsize}  
\tablecaption{PI SN Progenitor Models (Masses are in \Ms)\label{tab:t1}}
\tablehead{\colhead{run} & \colhead{$M_{init}$} & \colhead{$M_{He}$} & 
\colhead{$E_{ex}$ (B)}  & \colhead{$M_{Ni}$}  & \colhead{$Z$ (\Zs)}}
\startdata 
h150      &  150  &  109  &  42.0 &  9.2    &   0.14  \\
h200      &  200  &  130  &  65.0 &  39.2  &   0.14  \\
h500s0  &  500  &  94    &  3.7   &  2.2    &   0.43  \\ 
h500s4  &  500  &  76 & 10.4  &  0.18  &   0.43  \\
\enddata
\end{deluxetable}   

Shock breakout and expansion into the wind is modeled with the radiation hydrodynamics 
code RAGE \citep{rage}.  Our models include multi-species advection, energy deposition by 
radioactive decay of \Ni, grey flux-limited diffusion with LANL OPLIB 
opacities\footnote{http://aphysics2/www.t4.lanl.gov/cgi-bin/opacity/tops.pl} \citep{oplib}, and 
2-temperature physics.  We evolve mass fractions for the 15 even-numbered elements that 
are predominantly synthesized by PI SNe.  Both the self-gravity of the ejecta and the gravity
of any object formed at the center of the grid due to fallback are included in our simulations.  
Self-gravity is important when the shock is still deep inside the star because the potential 
energy of the ejecta is close to its kinetic, internal and radiation energies.  Failure to include 
it causes the shock to break out of the star with far higher energies and luminosities than it 
actually has.  Point-mass gravity is important in weak explosions in which much of the star 
may remain gravitationally bound and later fall back in on itself.     

\begin{figure}
\plotone{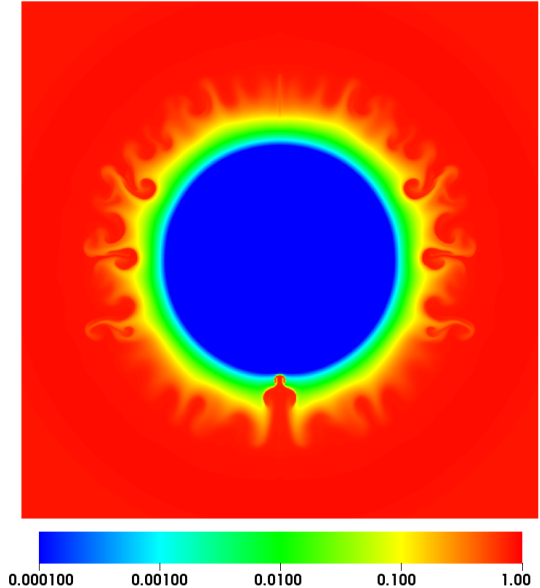}
\caption{Mixing in the oxygen shell (the site of explosive burning) in the 500 \Ms\ 0.1 \Zs\ PI 
SN; this image shows O mass fraction.  Mild instabilities form during explosive burning.  The 
shock is about to break through the surface of the He core on larger scales not shown in this 
image.  The minor mixing found in these PI SNe introduces only minor departures from 
spherical symmetry, justifying the use of 1D models for their light curves.} \vspace{0.1in}
\label{fig:mixing}
\end{figure}

We evolve the explosions on a 1D spherical coordinate grid in RAGE.  Several lines of 
evidence now show that internal mixing in PI SNe, which would break the spherical symmetry 
of the explosion, usually does not occur \citep[and is minor if it does;][]{jw11,chen11,chen14a,
chen14b}, in contrast to core-collapse explosions in which mixing is rampant prior to shock 
breakout \citep[e.g.,][]{fryer07,jet09b}.  We have run preliminary 2D CASTRO simulations 
\citep{Almgren2010} of the explosions in this study that show that if mixing occurs it is 
minimal, and mostly confined to the O shell (Fig.~\ref{fig:mixing}).  This is because the stars 
have shed their hydrogen envelopes prior to the explosion so breakout occurs from a He 
core, whose radius is just 1\% that of the original star.  Instabilities, if they form, have little 
time to grow before the shock reaches the surface.  This may not always be the case; stellar 
rotation has now been shown to induce dynamical instabilities into the explosion in 2D PI SN 
models \citep{cwc13}.

The RAGE mesh has 100,000 zones and a resolution of 1.2 $\times$ 10$^{6}$ cm for the 
h150 and h200 models and 2.0 and 6.0 $\times$ 10$^5$ cm for the h500s0 and h500s4 
models, respectively.  Outflow and reflecting boundary conditions for gas and radiation are 
imposed on the inner boundary, respectively, and outflow conditions are set on both flows
at the outer boundary. Because no compact objects are present in the Kepler profiles when 
they are ported to RAGE we set the point mass at the inner boundary to zero at the start of 
the run.  Any fallback to the center of the grid is then tallied and incorporated into the point 
mass, whose gravity can evolve during the simulation.  

We allocate 25\% of the grid to the ejecta at setup. The initial radius of the shock varies with 
the star but is typically 50 - 80\% that of the He core.  Up to two levels of adaptive mesh 
refinement (AMR) are applied in the initial interpolation of the profiles onto the grid and then 
during the simulation. To speed up the run and accommodate the expansion of the ejecta we 
resize the mesh by a factor of 2.5 every 10$^6$ time steps or when the radiation front has
crossed 90\% of the grid, whichever happens first.  The initial time step on which the new 
series evolves scales roughly as the ratio of the outer radii of the new and old grids.  We 
again apply up to 2 levels of refinement when mapping the explosion to a new grid and then
throughout the run.  The explosion is evolved out to 3 yr, until its luminosity has fallen below 
detection limits.  

Radiation energy densities are not explicitly evolved in the Kepler models so we initialize them 
in RAGE by 
\vspace{0.05in}
\begin{equation}
e_{\mathrm{rad}} = aT^4, \vspace{0.05in}
\end{equation}
where $a =$ 7.564 $\times$ 10$^{-15}$ erg cm$^{-3}$ K$^{-4}$ is the radiation constant 
and $T$ is the gas temperature.  Also, since the gas energy in Kepler includes contributions 
by ionization states of atoms, we unambiguously construct the specific internal energy from 
$T$ with
\vspace{0.05in}
\begin{equation}
e_{gas} = C_\mathrm{V}T, \vspace{0.05in}
\end{equation}
where $C_\mathrm{V }= $ 1.2472 $\times$ 10$^{8}$ erg gm$^{-1}$ K$^{-1}$ is the specific 
heat of the gas.  

\subsection{SPECTRUM}

To calculate a spectrum we first sample the RAGE radiation energy density profile from the 
outer boundary inward to find the position of the radiation front, which we take to be where 
$aT^4$ rises above 10$^{-4}$ erg/cm$^3$.  This corresponds to a radiation temperature
that is slightly higher than that of the surrounding wind (0.01 eV, or $\sim$ 116 K), so that the
outer layers of the front can be distinguished from the wind.  We then determine the radius 
of the $\tau = $ 40 surface by integrating the optical depth due to Thomson scattering inward 
from the outer boundary ($\kappa_{Th} =$ 0.288 for pristine H and He gas).  This yields the 
greatest depth in the flow from which most of the photons can escape because $\kappa_{Th}
$ is the minimum opacity that the photons would encounter.  

Densities, velocities, temperatures and species mass fractions are then extracted from the
RAGE profile and interpolated onto a 2D grid in $r$ and $\theta$ in SPECTRUM \citep{fet12},
whose inner and outer boundaries are zero and 2.0 $\times$10$^{17}$ cm, respectively.  We 
allocate 800 uniform zones in log radius from the center of the grid to the $\tau =$ 40 surface.  
The region from the $\tau =$ 40 surface and the radiation front is then partitioned into 6200 
uniform zones in radius. The wind between the radiation front and the outer boundary is 
divided into 500 uniform zones in log radius for a total of 7500 radial bins. 

The data in each radial bin is mass averaged so that SPECTRUM captures any sharp 
features in the RAGE profile. The SPECTRUM grid is uniformly discretized into 160 bins in 
$\mu =$ cos$\, \theta$ from -1 to 1.  Our choice of mesh ensures that regions of the flow 
from which photons can escape into the ISM are well resolved while those from which they 
cannot are only lightly sampled.  Calculating spectra in this manner allows us to determine 
how a dense wind shrouding the star absorbs radiation from the explosion, an effect not 
considered in earlier studies.  Light curves are constructed from 200 - 300 spectra.  The 
spectra are usually logarithmically spaced in time from shock breakout to 3 yr.

\subsection{Circumstellar Environment}

\begin{figure}
\begin{center}
\begin{tabular}{c}
\epsfig{file=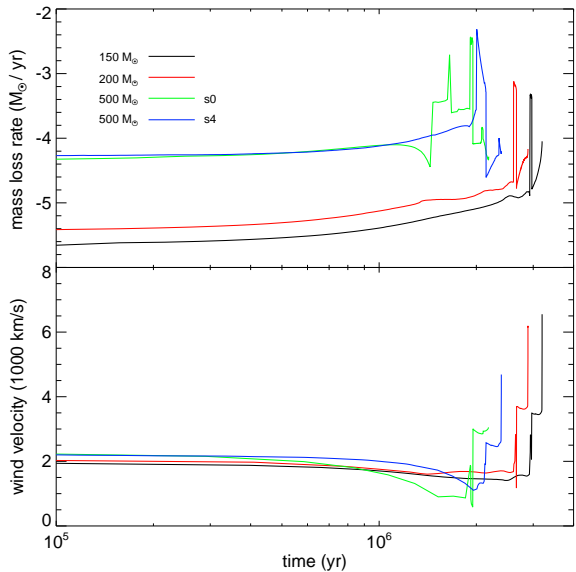,width=0.95\linewidth,clip=}  
\end{tabular}
\end{center}
\caption{Mass loss rates $\dot{m}$ (upper panel) and wind velocities $v_w$ (lower panel) 
for the four PI SN progenitors in our study.}
\label{fig:mdot}
\end{figure}

\begin{figure}
\begin{center}
\begin{tabular}{c}
\epsfig{file=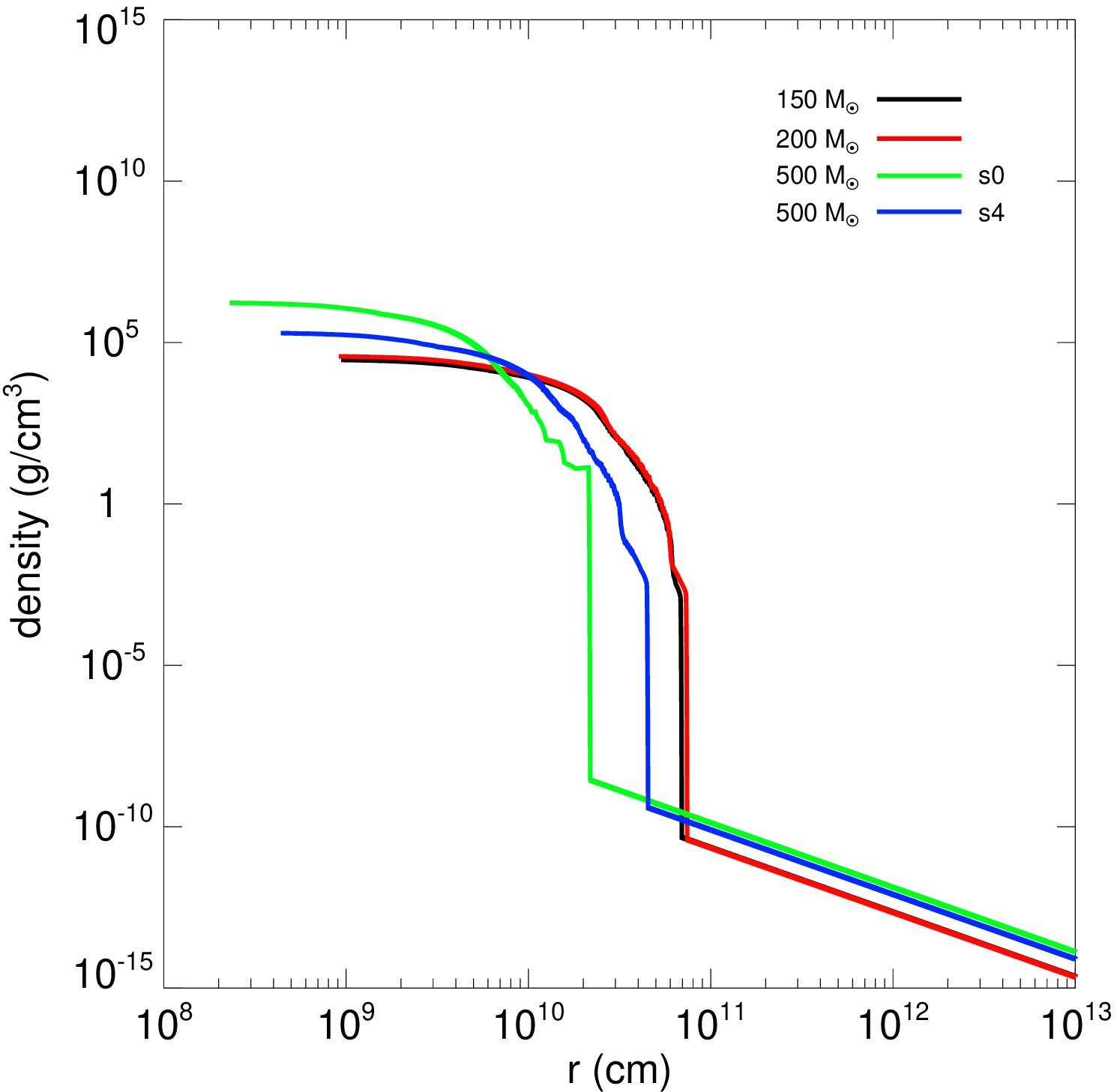,width=0.95\linewidth,clip=}  \\
\epsfig{file=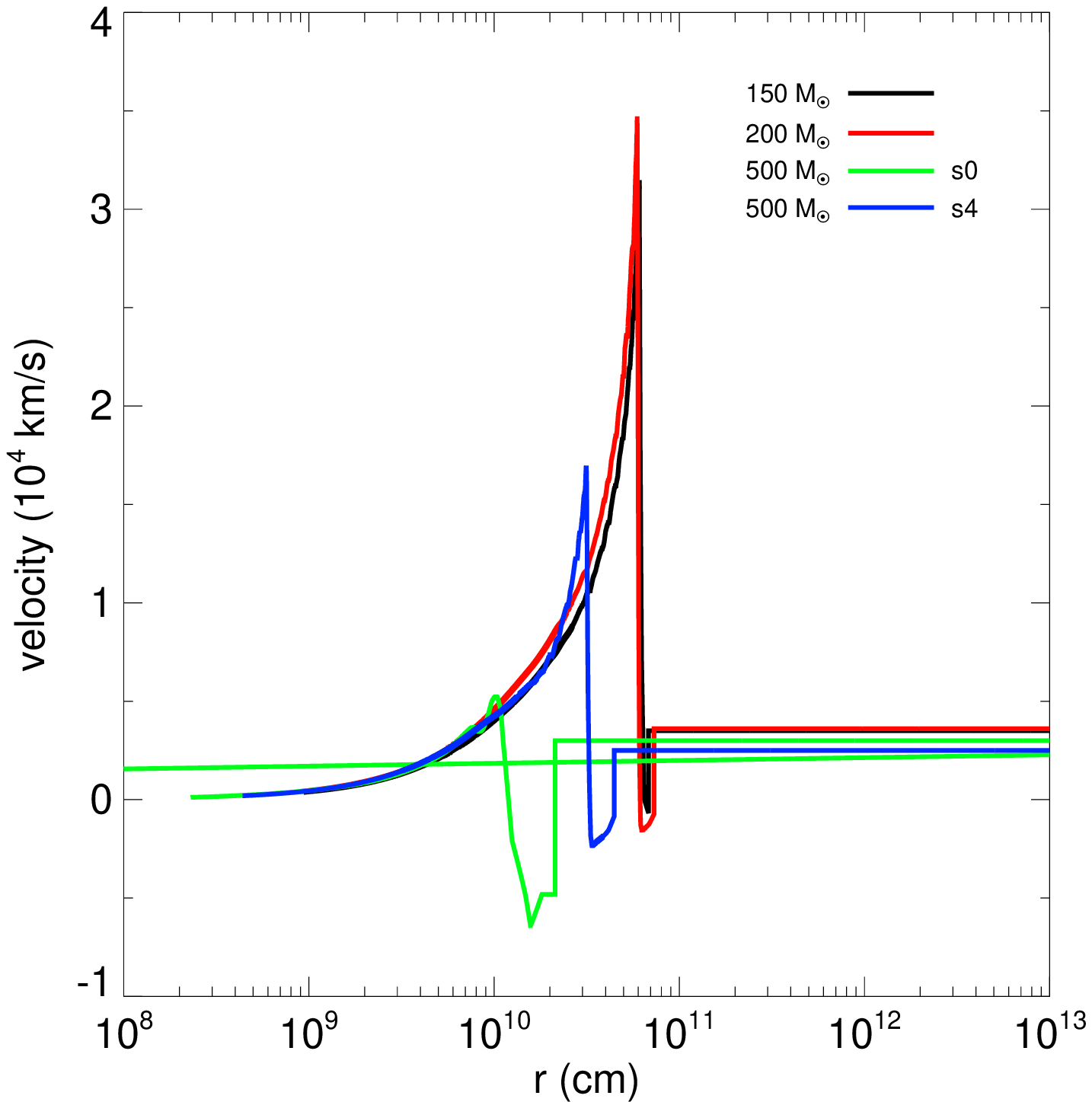,width=0.95\linewidth,clip=}  
\end{tabular}
\end{center}
\caption{Explosion and wind profiles for all four progenitors in dense wind envelopes just 
before breakout from the bare He core.  Top panel: densities; bottom panel: velocities.  
The radii of the cores vary from 10$^{10}$ - 10$^{11}$ cm.} \vspace{0.1in}
\label{fig:prof}
\end{figure} 

We show mass loss rates and wind velocities for the four progenitors in Fig.~\ref{fig:mdot}.  
The wind velocity, which is taken to be the escape velocity from the surface of the star and
varies from 1000 - 6000 km s$^{-1}$, is time-dependent because the radius of the star, 
which is also a product of our stellar evolution calculations, evolves over time.  The 
composition of the wind is primordial, 76\% H and 24\% He by mass.  The mass loss rates 
and wind speeds for the stars are fairly constant for their first 1 - 2 Myr but they all exhibit 
strong outbursts at the end of their lives.

\begin{deluxetable}{ccc}  
\tabletypesize{\scriptsize}  
\tablecaption{Mass Loss Rates and Wind Speeds\label{tab:t2}}
\tablehead{\colhead{$M_{\mathrm{init}}$ (\Ms)} & \colhead{$\dot{m}$ (\Ms\ yr$^{-1}$)} 
& \colhead{$v_{\mathrm{w}}$ (km s$^{-1}$)}}
\startdata 
150     &  10$^{-4.8}$  &  3500  \\
200     &  10$^{-4.8}$  &  3600  \\
500s0 &  10$^{-4.1}$  &  3000  \\
500s4 &  10$^{-4.4}$  &  2500  \\
\enddata
\end{deluxetable}   

As shown by \citet{met12a} with the ZEUS-MP code \citep{wn06}, outbursts and strong winds 
in principle can form quite complex structures around the star. The collision of the ejecta with 
such structures could strongly affect its luminosity.  But the SN expands to at most $\sim$ 0.3 
pc before becoming dim so it is only the mass loss in the few years following the final outburst 
of each star that sets the circumstellar profile out to this radius.  As shown in the bottom panel 
of Fig.~\ref{fig:mdot}, wind speeds for all four stars are relatively constant in the 10 - 100 yr 
after the outburst.  Mass loss rates immediately after the ejections are also relatively constant 
over these times.  The stellar envelope is therefore well-approximated by the usual power-law 
density profile
\begin{equation}
\rho \, = \, \frac{\dot{m}}{4 \pi r^2 v_{\mathrm{w}}}, \vspace{0.05in} \label{eq:wind}
\end{equation}
where $v_{\mathrm{w}}$ and $\dot{m}$ are the wind speed and mass loss rate just after the 
ejection, respectively.  We summarize $v_{\mathrm{w}}$ and $\dot{m}$ in Table~\ref{tab:t2}.  
To soften the density drop at the surface of the star (and thus prevent numerical instabilities 
at shock breakout) we bridge the star and the wind by an $r^{-25}$ density profile. We show 
density and velocity profiles for the shock, the star, and the surrounding wind in 
Fig.~\ref{fig:prof}.

We consider two limiting cases for the supernovae: explosions in the dense winds just 
described and explosions in diffuse envelopes in which the outburst clears gas from the 
vicinity of the star.  In this latter case we use Eq.~\ref{eq:wind} for the wind with $v_
\mathrm{w} =$ 1000 km s$^{-1}$ and an $\dot{m}$ chosen to guarantee that $\rho_{
\mathrm{w}} \sim$ 2 - 3 $\times$ 10$^{-18}$ g cm$^{-3}$ at the bottom of the density 
bridge from the surface of the star.  We examine explosions in heavy winds first, taking 
the h150 PI SN as a fiducial case.  

\section{Explosions in Dense Envelopes}

\begin{figure}
\begin{center}
\begin{tabular}{c}
\epsfig{file=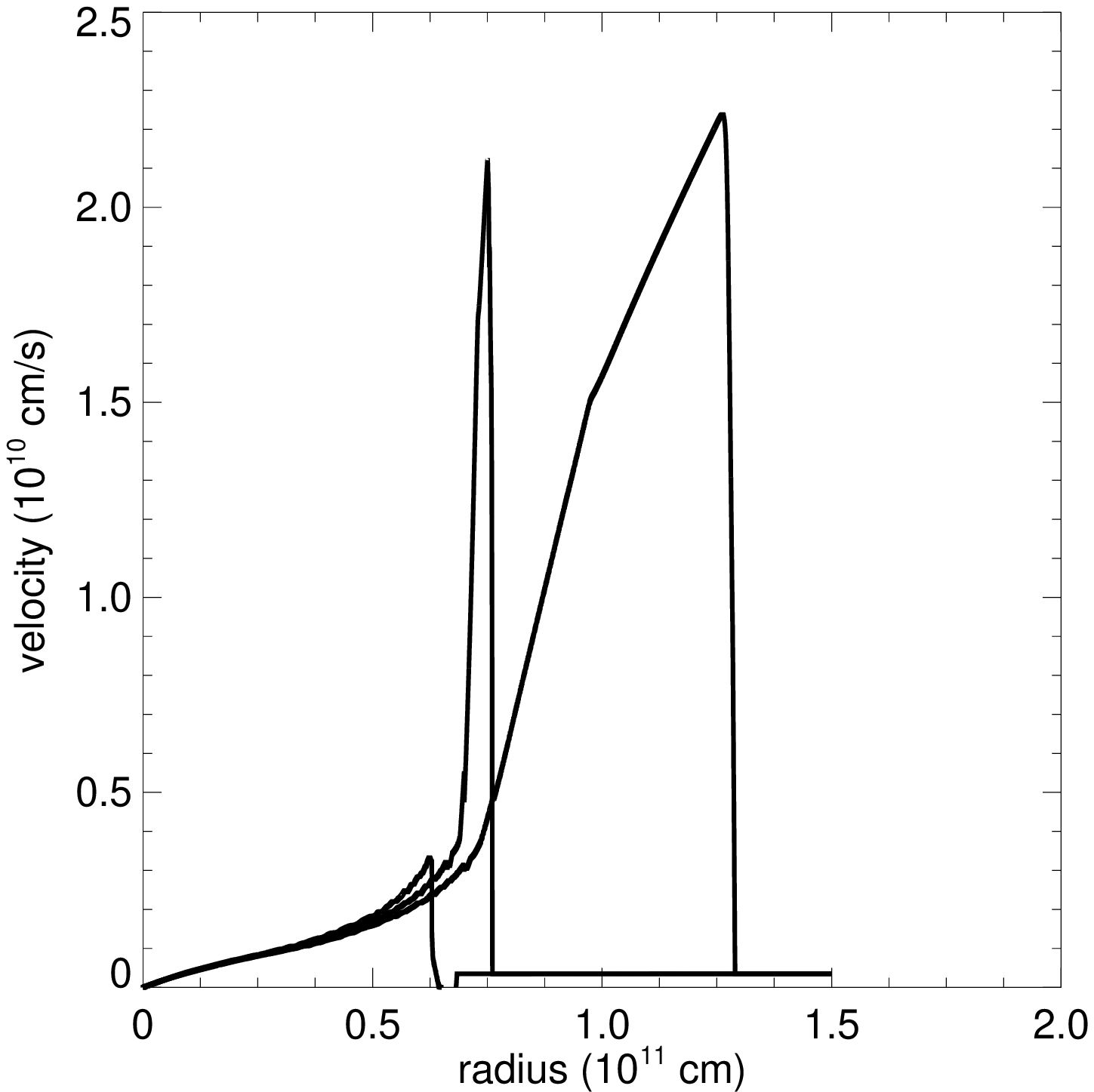,width=0.85\linewidth,clip=} \\ 
\epsfig{file=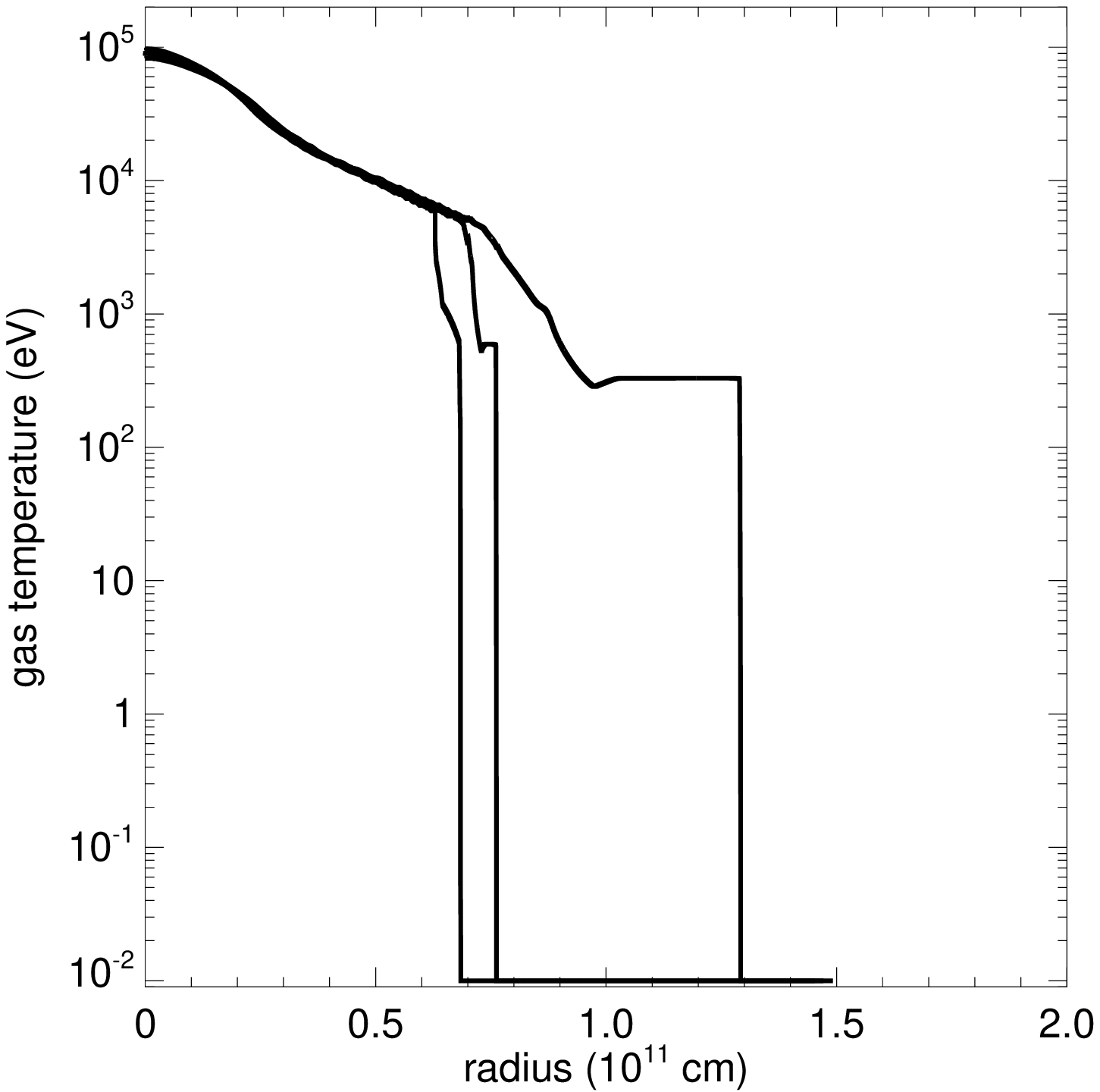,width=0.85\linewidth,clip=} \\
\epsfig{file=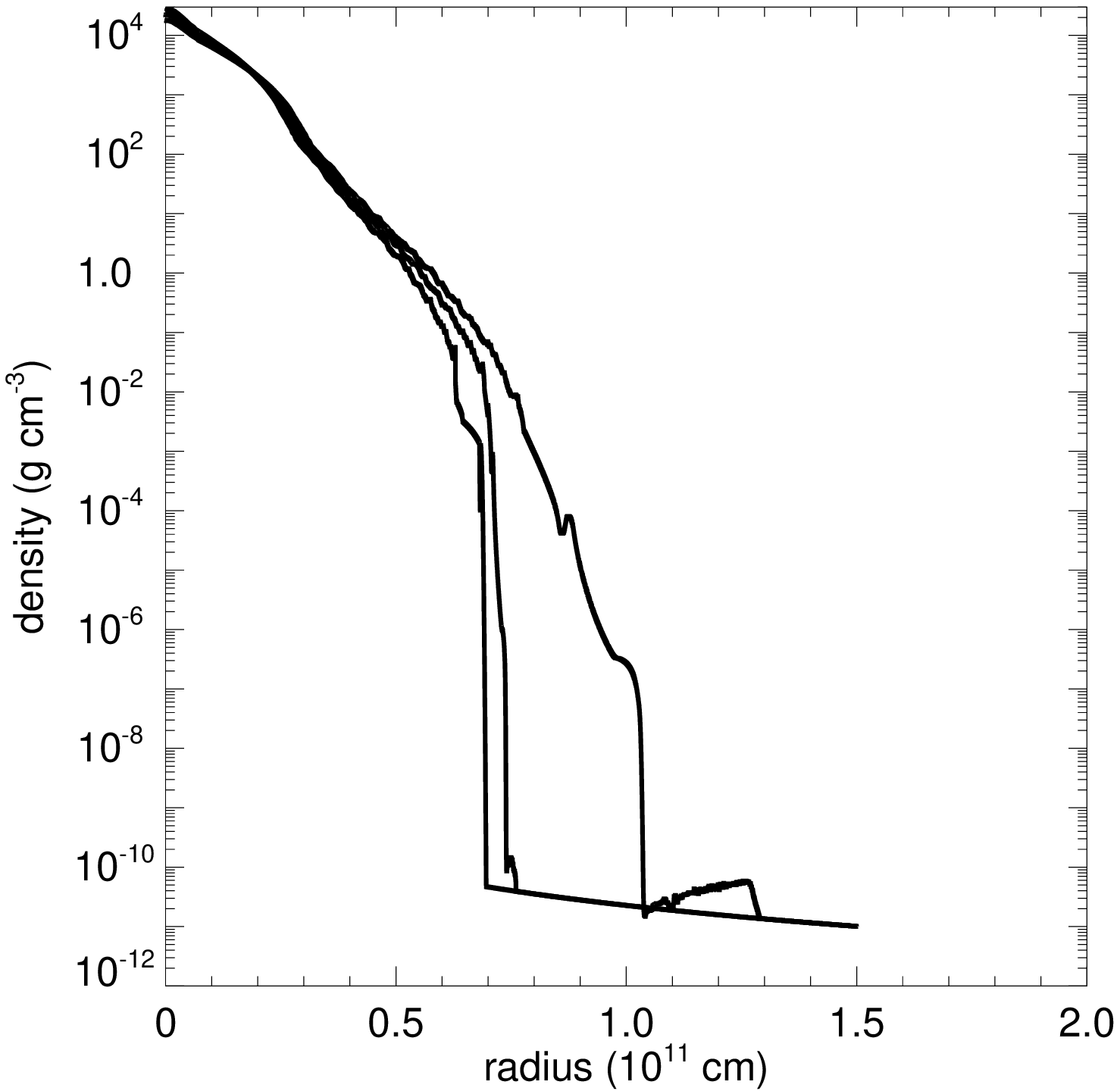,width=0.85\linewidth,clip=} 
\end{tabular}
\end{center}
\caption{Shock breakout for the h150 PI SN in a dense envelope. Top:  velocities; center: 
temperatures; bottom:  densities.  From left to right the times are 0.46 s, 2.2 s, and 3.9 s.}
\label{fig:sbo1}
\end{figure}

\subsection{Shock Breakout}

\begin{figure*}
\plottwo{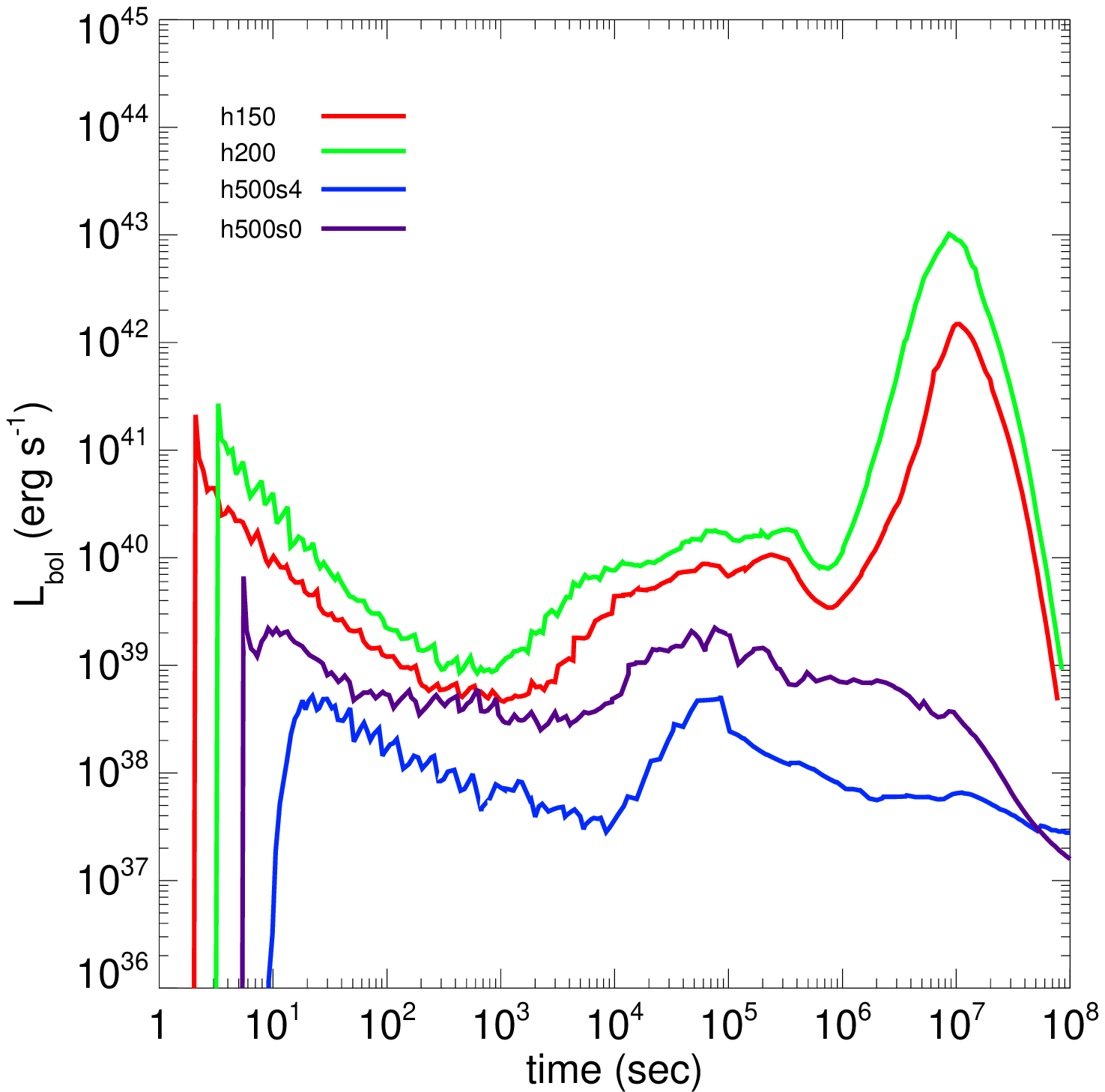}{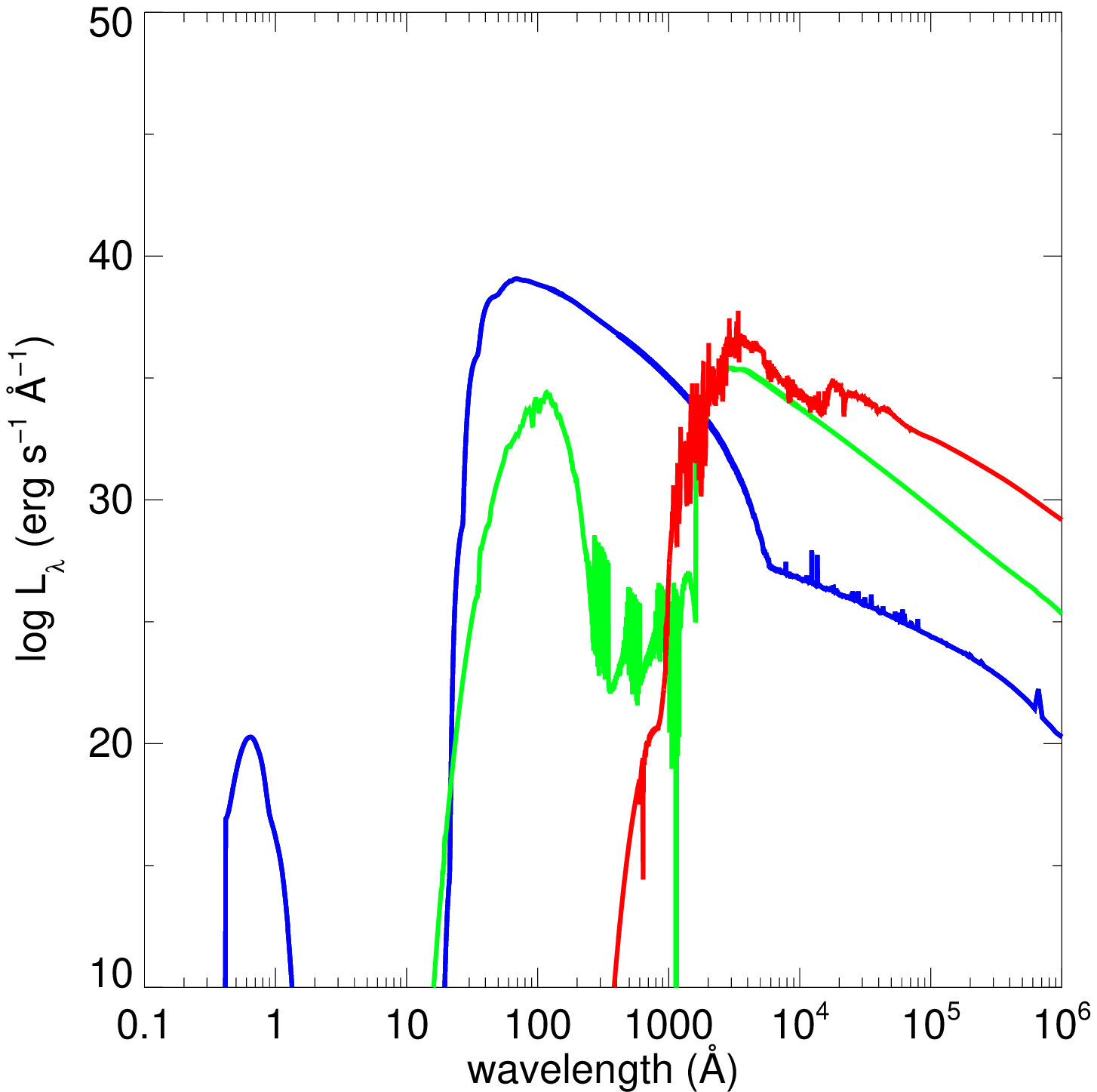}
\caption{Left panel:  bolometric luminosities for all 4 PI SNe in dense winds.  Right panel: 
spectral evolution of the h150 PI SN.  Blue:  shock breakout from the star at 2.2 s.  Green:  
partial quenching of the spectrum at 234 s due to recombination of the wind after being 
briefly ionized by the breakout pulse.  Red:  radiation breakout from the envelope at 1.40 
$\times$ 10$^5$ s.} \vspace{0.1in}
\label{fig:bLC1}
\end{figure*} 

\begin{figure*}
\plottwo{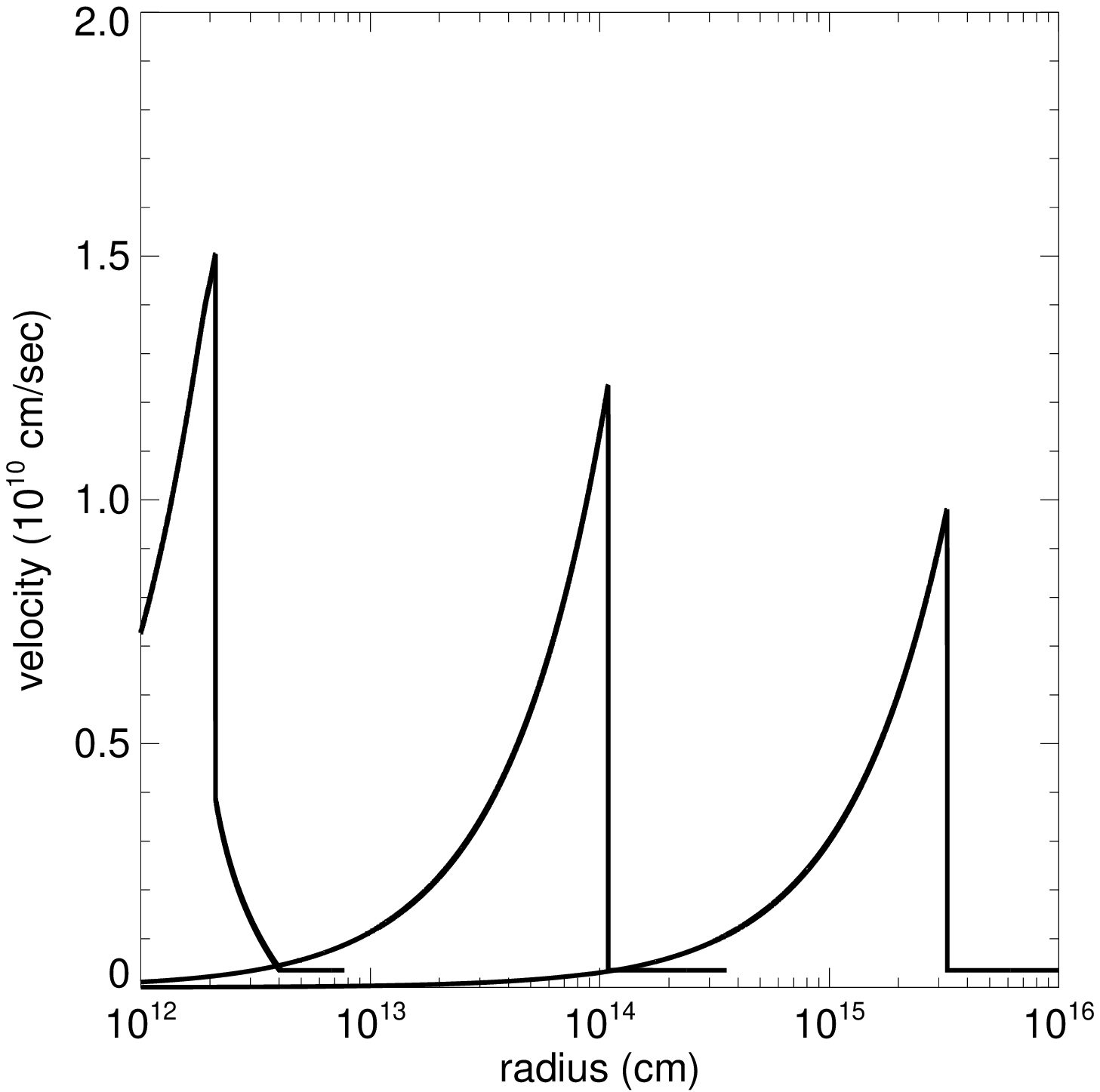}{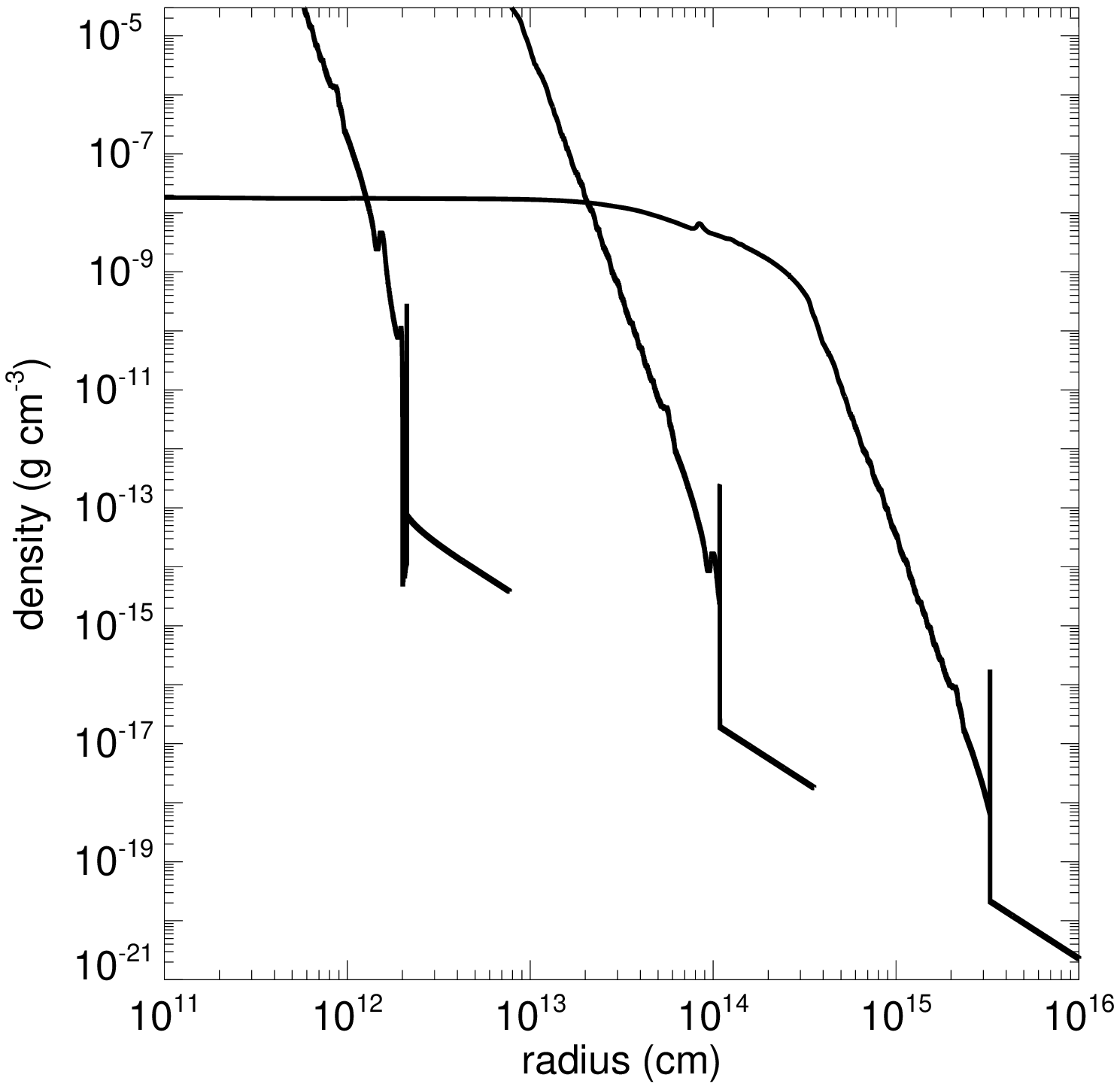}
\caption{Evolution of the h150 PI SN at intermediate times.  Left panel:  velocities, from
left to right, at 133 s, 8792 s, and 3.32 $\times$ 10$^5$ s.  Right panel: densities at the
same times.} \vspace{0.1in}
\label{fig:int1}
\end{figure*} 


Velocity, temperature and density plots for shock breakout from the surface of the star 
are shown in Fig.~\ref{fig:sbo1}. Prior to breakout the shock is not visible to an external 
observer because of scattering by free electrons in the upper layers of the star.  When 
the shock breaks through the surface of the star it abruptly accelerates in the steep 
density gradient there, as shown in the velocities at 0.46 s and 2.2 s.  At the same time, 
photons that were previously trapped in the shock and simply advected along by the 
flow now break free of it, and a radiation pulse propagates into the wind.  The radiation 
front is visible as the flat plateau in temperature ahead of the shock at 2.2 s and 3.9 s.  
The temperature of the plateau ($\sim$ 500 eV at 2.2 s and 300 ev at 3.9 s) is the 
temperature to which the radiation heats the wind, not the shock itself, which is much 
hotter ($\sim$ 2 keV). The temperature of the plateau falls over time because the shock 
expands and cools and its spectral peak shifts to lower energies.

The breakout pulse is usually thought to be approximately the light crossing time of the 
star (here, a few seconds for the compact He core), but in reality it is somewhat longer
because the photons do not instantly decouple from the shock. Instead, the pulse blows 
off the outermost layers of the star at very high velocities, as shown at 2.2 and 3.9 s.  At
3.9 s the shock is visible in the density profile at 9.5 $\times$ 10$^{10}$ cm.  The wispy 
outer layers of the star have been driven ahead of the shock by the radiation to 1.4 
$\times$ 10$^{11}$ cm.  The shock drives this radiative precursor into the wind until it 
cools, dims, and can no longer do so, as we show later.  Radiation/matter coupling is 
especially strong in breakout from compact cores because the temperature of the shock 
is so high, twice that of breakout in the z-series PI SNe in \citet{wet12b}, because the 
shock has done less $PdV$ work on its surroundings.  We also note that because the 
opacities are frequency dependent, photons break free of the flow at different times at 
different wavelengths.  This effect also lengthens the pulse in time \citep{bay14}.

In most SNe the breakout transient is composed primarily of x-rays and hard UV, and 
in Type Ia and II explosions its luminosity is $\sim$ 10$^{42}$ - 10$^{45}$ erg s$^{-1}$.  
But the dense wind quenches most of the transient in our first fours explosions, limiting 
its initial brightness to $\sim$ 10$^{39}$ - 10$^{41}$ erg s$^{-1}$ as shown in the left 
panel of Fig.~\ref{fig:bLC1}.  Here, the initial flux is mostly UV and optical, with a few 
hard x-rays as shown in the spectrum at 2.2 s in the right panel of Fig.~\ref{fig:bLC1}.  
The radiation pulse briefly ionizes part of the envelope at breakout which then rapidly 
recombines as the transient dies down, partially shuttering subsequent radiation from 
the fireball from $\sim$ 2 s to 10$^3$ s as shown in the spectra in the right panel of 
Fig.~\ref{fig:bLC1}. 

At 234 s the only emission from the fireball that escapes the dense shroud is optical, 
IR, and some UV.  But the luminosity again begins to rise as the shock grows in radius 
and more of its photons escape the wind, as shown from 10$^3$ s to 10$^5$ s.  As 
the wind beyond the shock becomes thin the UV emission rises again, as shown in the 
spectrum at 1.40 $\times$ 10$^5$ s.  Thus, there is an initial, partial breakout of 
radiation from the envelope at shock breakout and then a second, more protracted 
breakout as the wind becomes optically thin to the shock.  Note that peak initial 
breakout luminosities rise with explosion energy, and that breakout happens later in
more massive progenitors with lower explosion energies because it takes longer for 
the shock to reach the surface of the star.

\begin{figure*}
\epsscale{2.3}
\plottwo{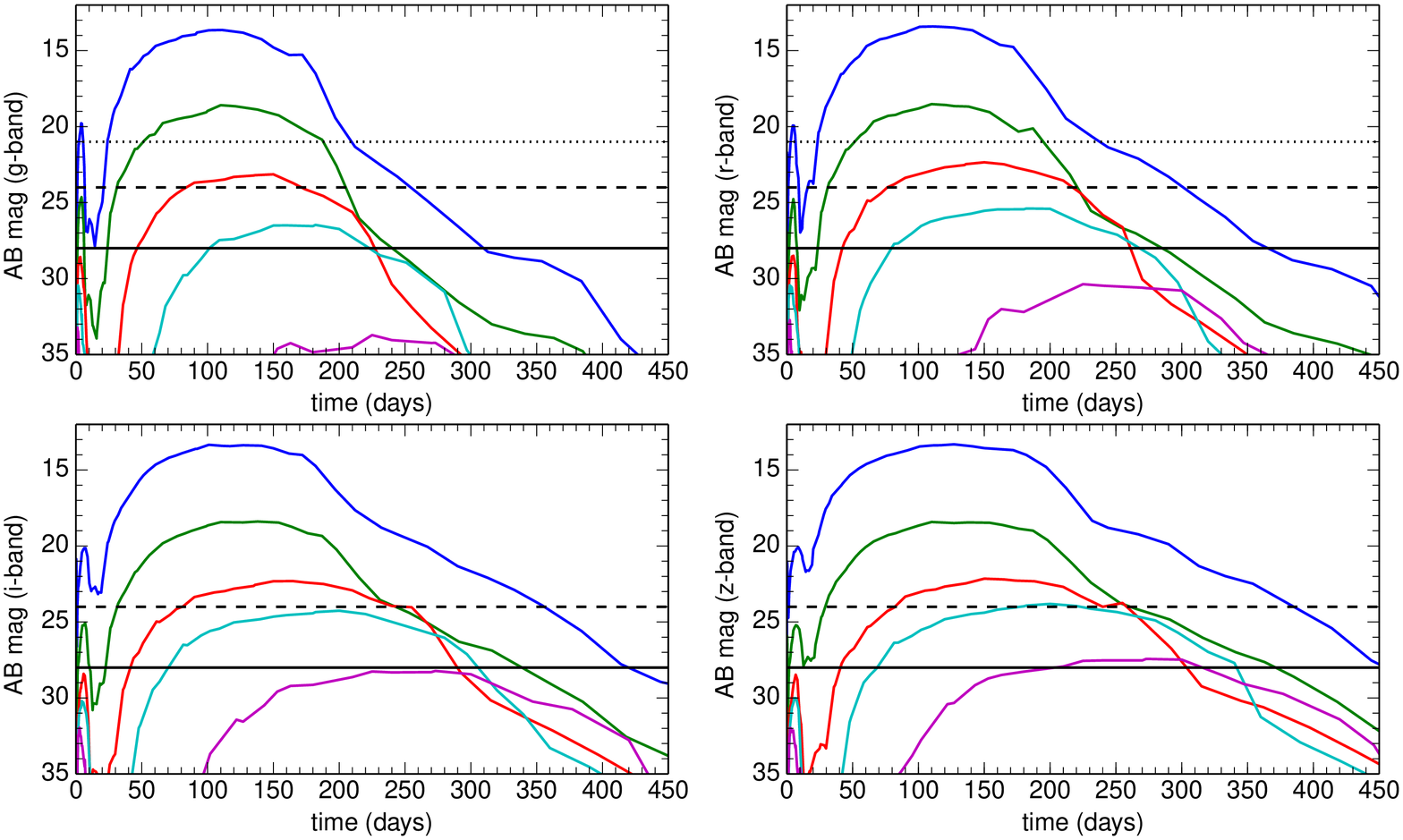}{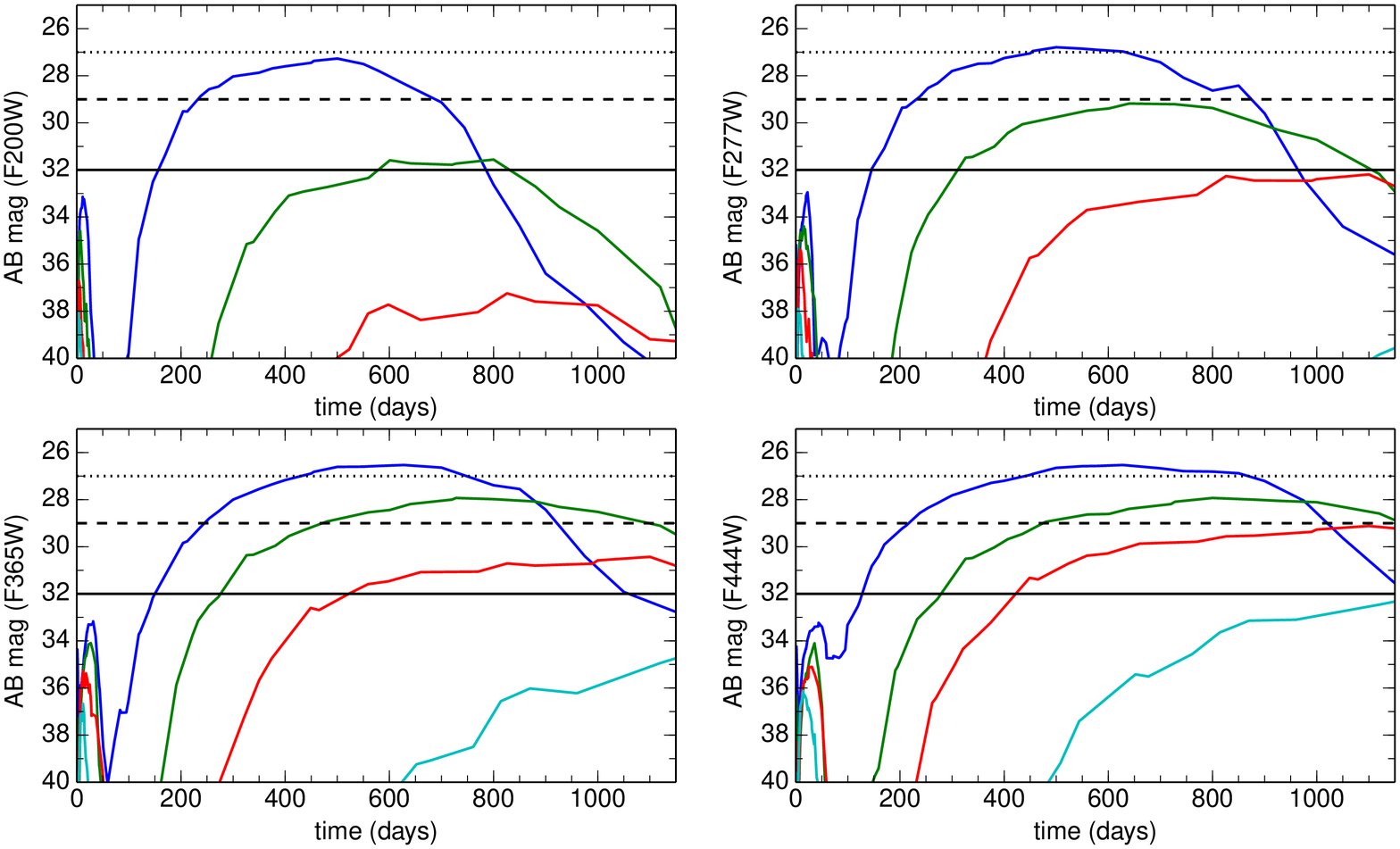}
\caption{Optical and NIR light curves for the h200 PI SN in a high-density envelope.  
Upper 4 panels:  $z =$ 0.01 (blue), $z =$ 0.1 (green), $z =$ 0.5 (red), $z =$ 1 (teal), 
and $z =$ 2 (purple).  Dotted line:  PTF detection limit; dashed line:  Pan-STARRS 
detection limit; solid line:  LSST detection limit.  Lower 4 panels:  $z =$ 4 ({\it dark 
blue}), 7 ({\it green}), 10 ({\it red}) and 15 ({\it teal}).  The horizontal dotted, dashed 
and solid lines are photometry limits for WFIRST, WFIRST with spectrum stacking 
and {\it JWST}, respectively.  The wavelength of each filter can be read from its 
name; for example, the F277W filter is centered at 2.77 $\mu$m, and so forth.} 
\vspace{0.1in}
\label{fig:LC1}
\end{figure*}

\subsection{Intermediate / Later Stages of the SN}

As the SN expands it plows up the wind and briefly continues to drive the radiative 
precursor ahead of it.  But by 133 s this precursor has collapsed back to the shock,
which has cooled to the point that it can no longer sustain the precursor as shown
in Fig.~\ref{fig:int1}.  As the fireball plows up more wind a reverse shock begins to
form, which is visible as the small density bump just behind the shock at 9 $\times$ 
10$^{13}$ cm.  This shock partially detaches from the forward shock and begins to 
step back through the flow in the frame of the ejecta.  As it does, the gas that is 
heated by the reverse shock radiatively cools, and the shock loses pressure support 
and retreats back into the forward shock. As the forward shock continues to plow up 
the envelope the cycle repeats.  This is the classic radiative instability that has been 
well studied by \citet{chev82} and \citet{imam84}, and the fluctuations in the 
postshock gas temperatures due to the oscillation of the revere shock are what
cause the ripples in the bolometric luminosities from 10$^4$ - 10$^5$ s.  

\begin{figure*}
\epsscale{1.1}
\plotone{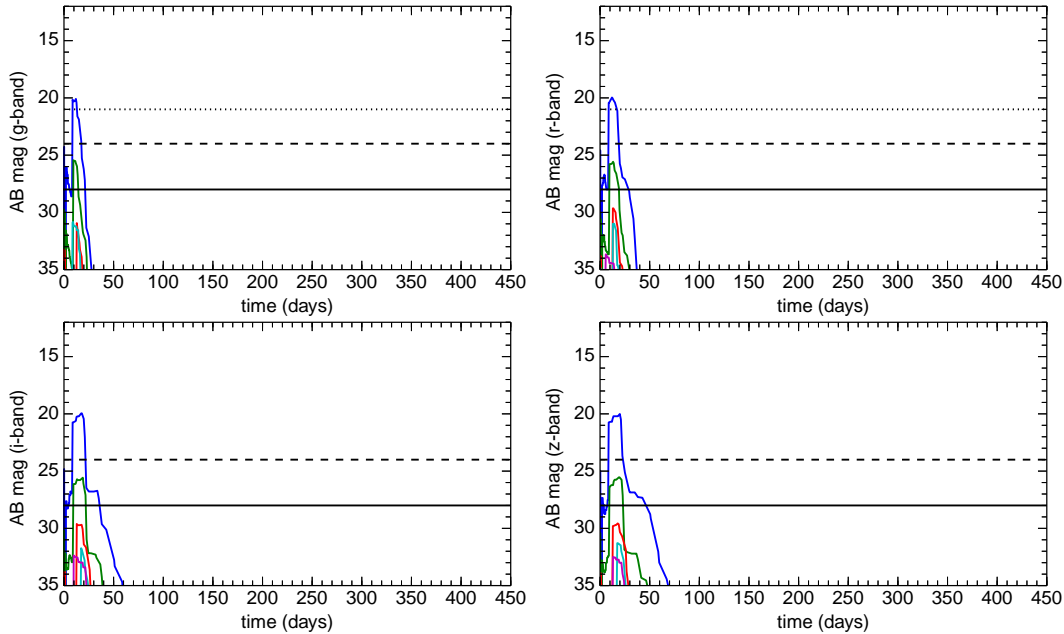}
\caption{Optical light curves for the h500s0 SN in a high-density envelope:  $z =$ 
0.01 (blue), $z =$ 0.1 (green), $z =$ 0.5 (red), $z =$ 1 (teal), and $z =$ 2 (purple).  
Dotted line:  PTF detection limit; dashed line:  Pan-STARRS detection limit; solid 
line:  LSST detection limit.} \vspace{0.1in}
\label{fig:LC2}
\end{figure*}

In a multidimensional simulation, the interface between the ejecta and the envelope 
in which the radiative instability occurs would likely be broken up by Rayleigh-Taylor 
(RT) instabilities, and this might reduce or remove altogether the fluctuations in the 
light curve over this time interval.  In principle, such instabilities could also disrupt 
the structures emitting most of the radiation from the shock and alter the luminosity 
of the explosion. But because the ripples themselves have small amplitudes, it is not 
likely that the mean luminosity would change much in their absence.  After 10$^5$ s
the expansion of the ejecta is nearly self-similar.  

By a few 10$^6$ s the SN ejecta has expanded to the point where its photosphere 
has descended down to the warm \Ni\ layer, and the h150 and h200 light curves 
exhibit the classic rebrightening associated with radioactive decay in PI SNe that 
synthesize large masses of \Ni.  But \Ni\ rebrightening happens at earlier times in 
our explosions than in non-rotating Pop III PI SNe, whose progenitors retain their 
mass.  This is due to the lower intervening masses between the \Ni\ layer and 
surrounding wind, and hence shorter radiation diffusion times. The h200 SN has 
higher peak luminosities than h150 because it makes four times the \Ni, 39 \Ms\ 
versus 9 \Ms.  The h500 light curves do not exhibit prominent rebrightening 
because they create far less \Ni, and in the s0 explosion it all falls back into a 
central BH, as we discuss later.  We note that even though the h500 PI SNe do 
not rebrighten, they do exhibit extended, dim emission due to shock heating as 
they propagate through the envelope, which lasts for nearly three years in the
h500s4 explosion.  

\subsection{Optical/NIR Light Curves}

NIR surveys are required to detect SNe prior to the end of reionization ($z \gtrsim
$ 6) because flux blueward of the Lyman limit at higher redshifts is absorbed by 
the neutral IGM.  This also restricts detections of such events in the optical to $z < 
$ 6. All-sky surveys offer the best prospects for detecting large numbers of high-$z
$ SNe because their large survey areas can compensate for low star formation 
rates (SFRs) at early epochs \citep[e.g., Fig.~3 of][]{wet13c}. But even 30-m class 
telescopes with narrow fields such as {\it JWST}, the Thirty-Meter Telescope 
(TMT), the {\it Giant Magellan Telescope} ({\it GMT}) and the European Extremely 
Large Telescope (E-ELT) are still expected to detect Pop III SNe in appreciable 
numbers \citep{hum12}.  We now examine detection limits in redshift for our PI 
SNe in the NIR for SNe at $z >$ 6 and in the optical for events below this redshift. 

We show g, r, i and z band light curves for the h200 and h500s0 explosions in 
dense envelopes in Figures~\ref{fig:LC1} and \ref{fig:LC2}, together with 
sensitivity limits for PTF, Pan-STARRS and LSST.  The h200 light curves all 
exhibit a initial, short-lived peak corresponding to the post-breakout expansion 
and cooling of the fireball followed by a second brighter and much longer peak 
due to \Ni\ rebrightening.  This explosion will be visible to PTF out to $z \sim$ 
0.1, to Pan-STARRS out to $z \sim$ 0.5 - 1, and to LSST out to $z \sim$ 2.  
On the other hand, the h500s0 PI SN is a much dimmer event, only being 
marginally visible to PTF and Pan-STARRS out to $z \sim$ 0.01 and to LSST 
out to $z \sim$ 0.1.  These two fiducial cases underscore the fact that PI SNe 
of non-zero metallicity stars can be either very bright or dim events, and that in 
the latter case some of these explosions could be hidden in a wide range of SN 
classes.

NIR light curves at 2 - 4 $\mu$m for the h200 run for $z =$ 4, 7, 10, and 15 are 
shown in Fig.~\ref{fig:LC2}.  We find that this explosion will be visible to WFIRST 
out to $z \sim$ 4 - 10 and to {\it JWST} out to $z \sim$ 10 - 15, the era of first 
galaxy formation. Some numerical models predict that most massive stars at 
this latter epoch will be contaminated by metals from the first generations of 
stars, so these events could be used to probe the stellar populations of the first
galaxies.  On the other hand, we find that peak AB magnitudes in the NIR for the 
more massive h500s0 and h500s4 explosions are well below even {\it JWST} 
detection limits at $z >$ 4.

\section{Explosions in Diffuse Envelopes}

\subsection{Shock Breakout}

\begin{figure}
\begin{center}
\begin{tabular}{c}
\epsfig{file=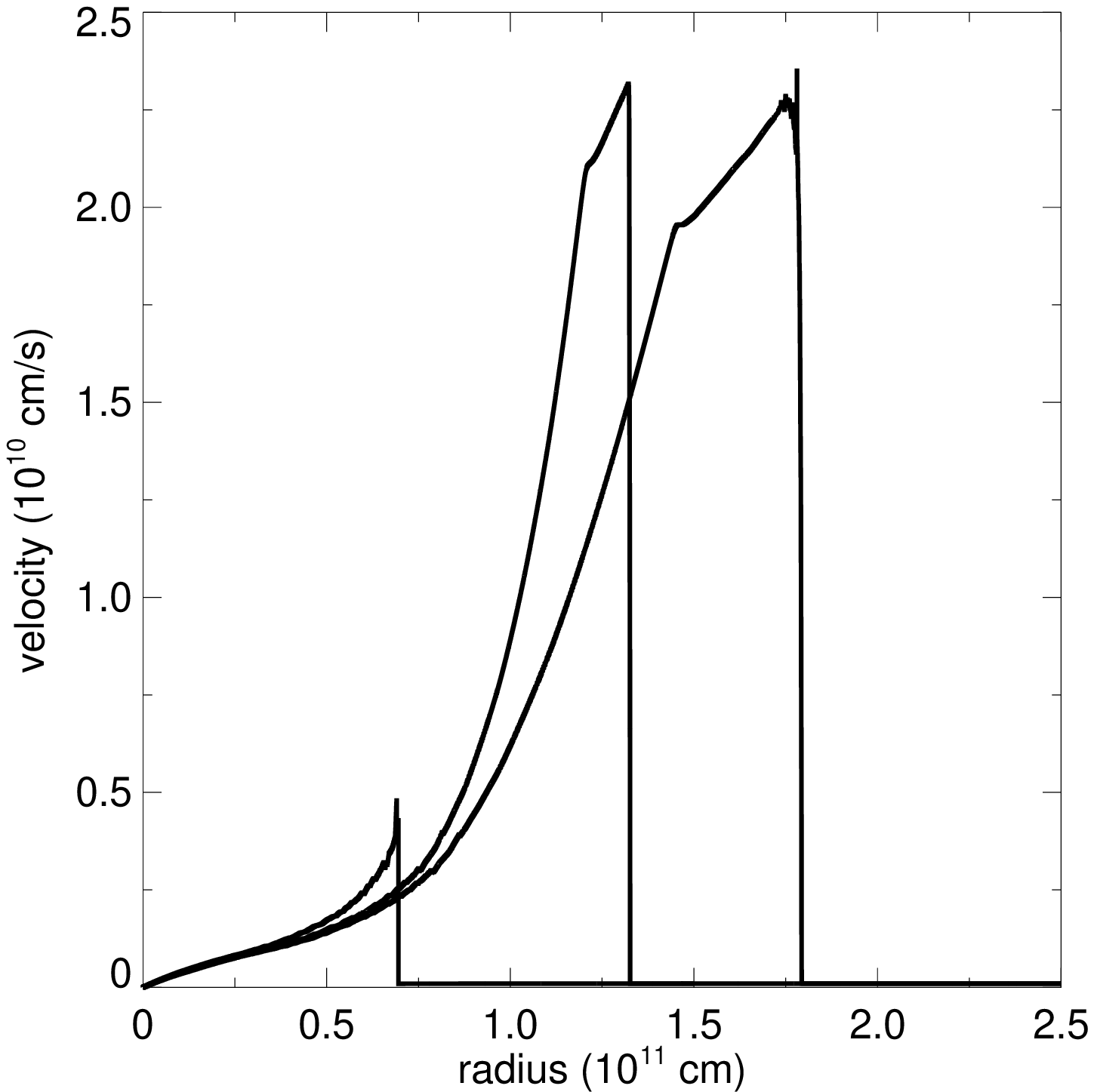,width=0.85\linewidth,clip=} \\ 
\epsfig{file=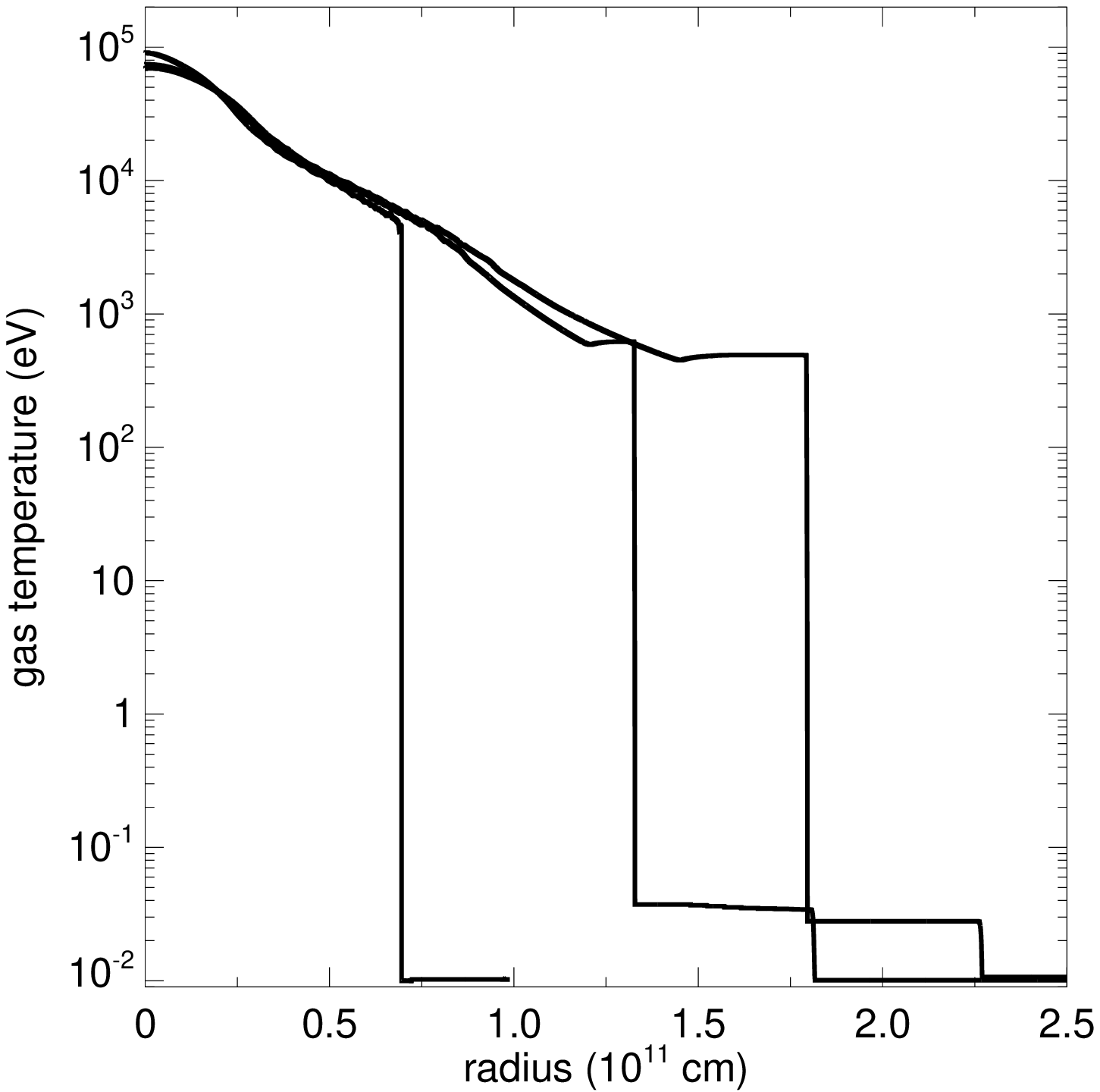,width=0.85\linewidth,clip=} \\
\epsfig{file=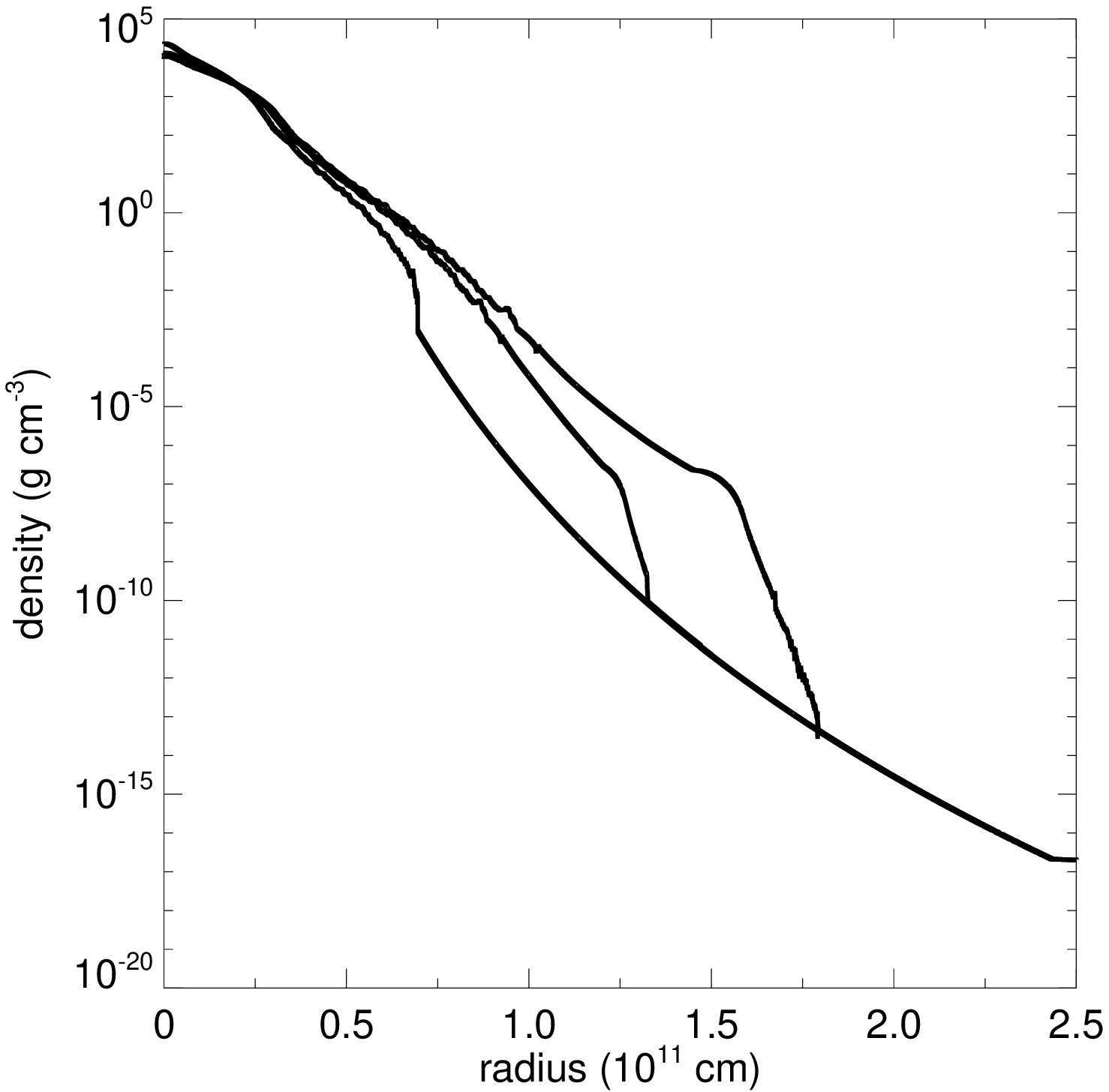,width=0.85\linewidth,clip=} 
\end{tabular}
\end{center}
\caption{Shock breakout for the h150 PI SN in a diffuse envelope. Top:  velocities; center: 
temperatures; bottom:  densities.  From left to right the times are 2.01 s, 6.18 s, and 7.69 
s.}
\label{fig:sbo2}
\end{figure}

Velocity, temperature and density profiles for shock breakout in a diffuse wind are shown
for the h150 PI SN in Fig.~\ref{fig:sbo2}.  We first note that ambient densities in this model
are five orders of magnitude lower than those in the dense envelopes.  The initial effect of 
the much lower density is that the shock heats the surrounding gas to higher temperatures 
after breakout, nearly 1 keV, and it remains hotter for longer times.  The shock cools more 
rapidly in dense winds due to the $PdV$ work it performs on its surroundings.  The shock
therefore has a harder spectrum when it breaks out into diffuse envelopes and, as we show 
later, far higher breakout luminosities because the wind at the base of the bridge is basically 
optically thin.  In these envelopes, shock breakout and radiation breakout occur at the same 
time and the shock reaches peak luminosity before reaching the bottom of the bridge.  
Blowoff and the formation of a radiative precursor is again visible in the broken slope of the 
velocities near their peak at 6.18 and 7.69 s.

\begin{figure*}
\begin{center}
\begin{tabular}{cc}
\epsfig{file=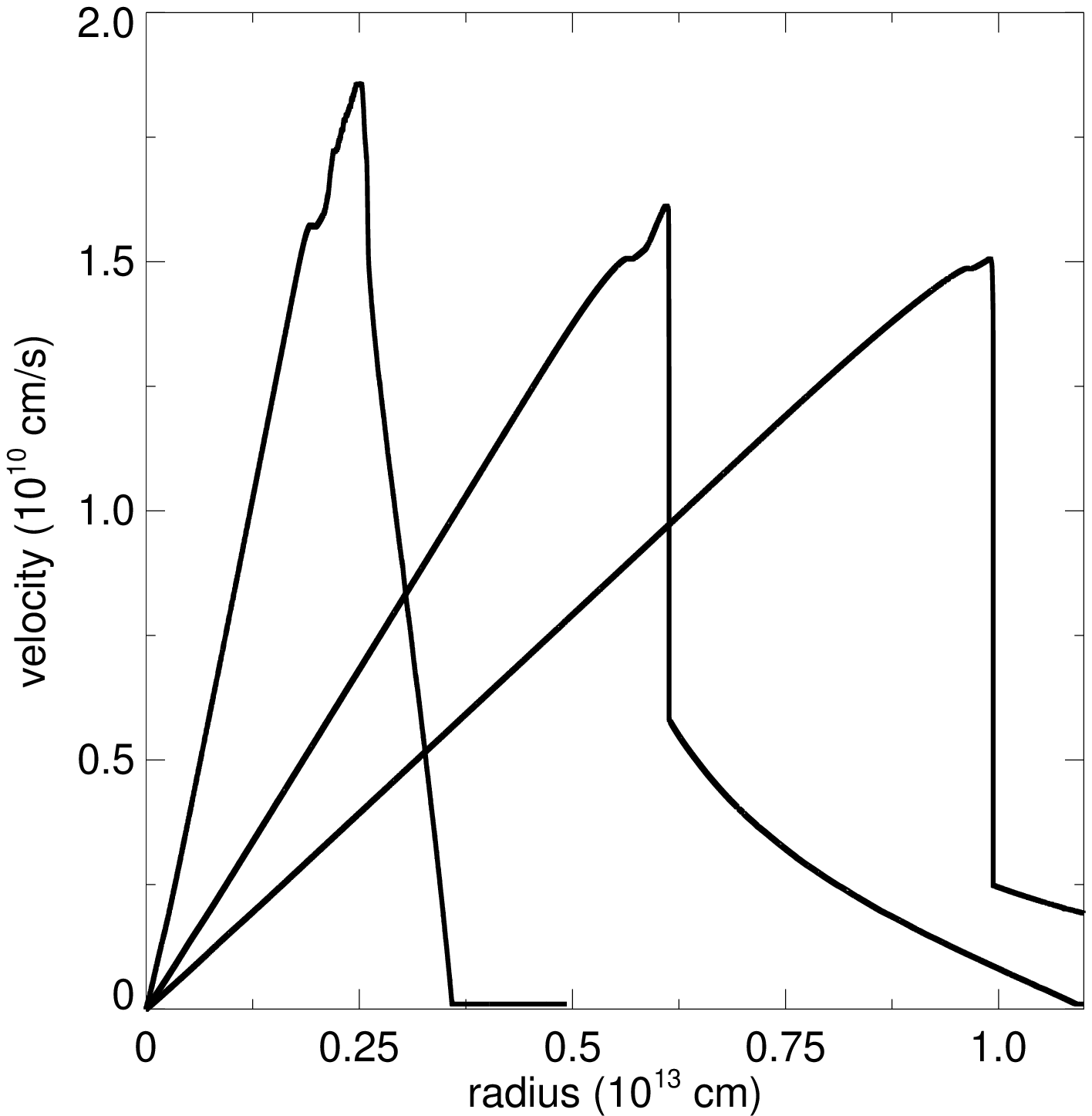,width=0.4\linewidth,clip=}  &
\epsfig{file=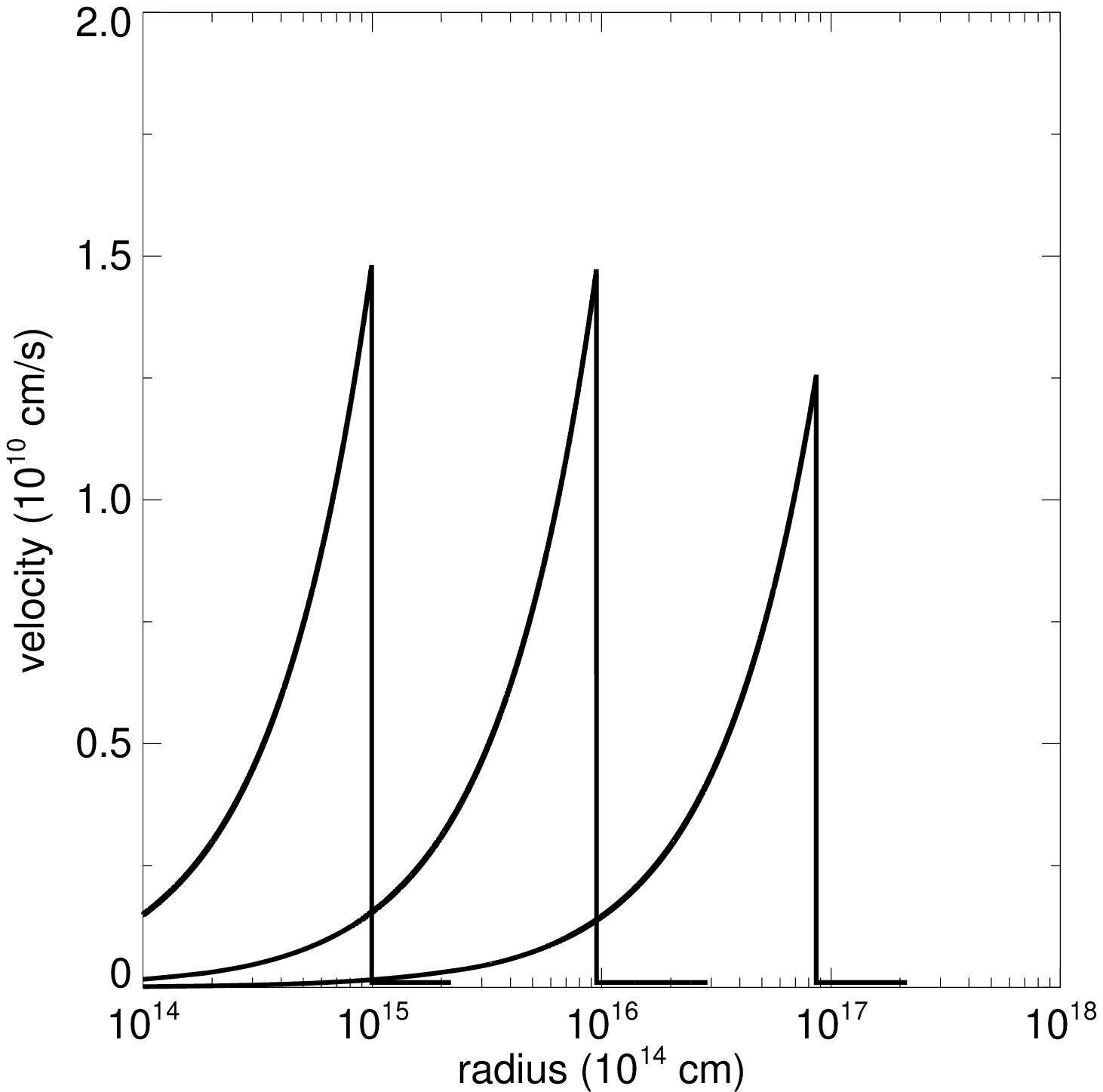,width=0.4\linewidth,clip=}  \\
\epsfig{file=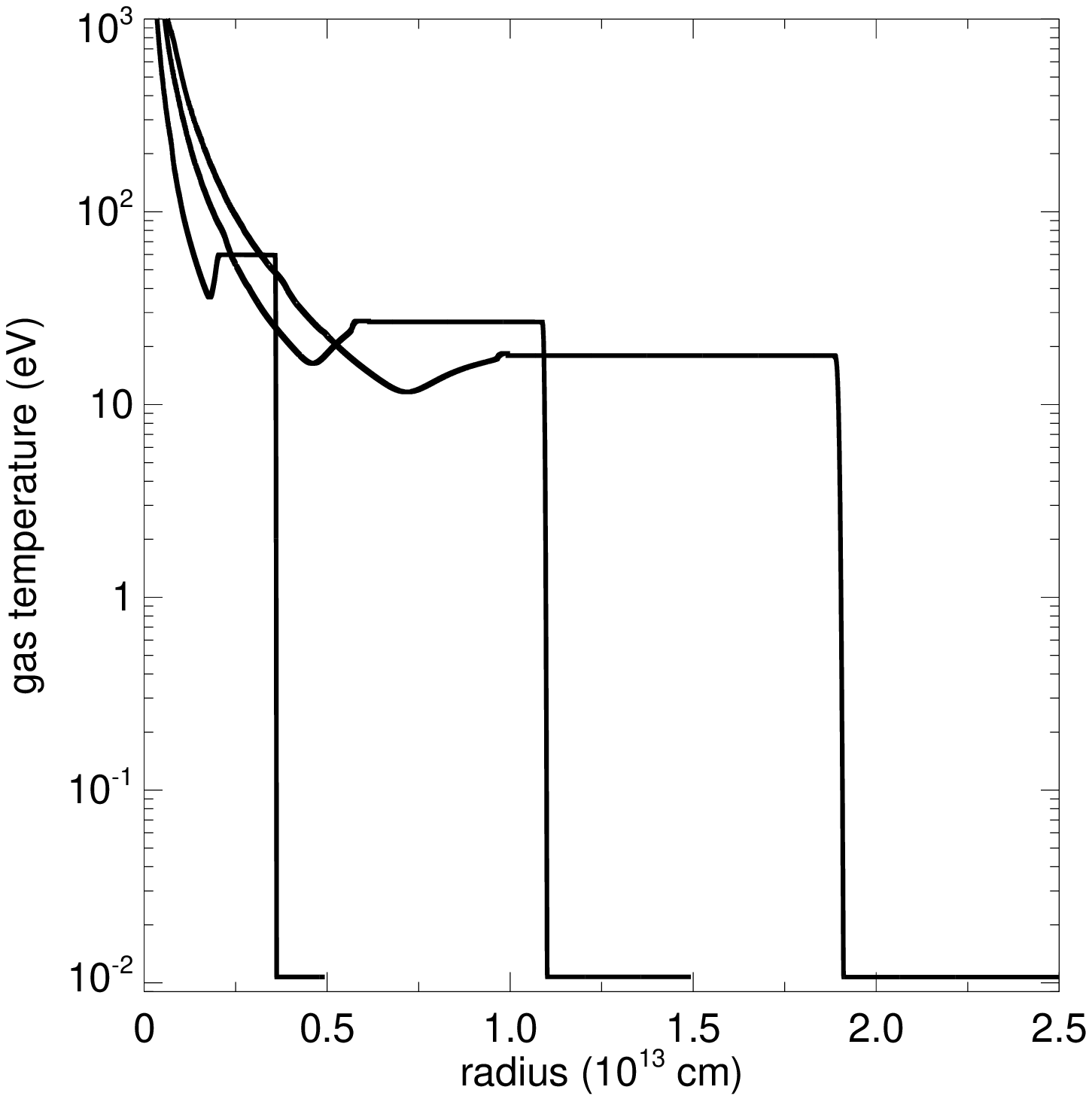,width=0.4\linewidth,clip=}  & 
\epsfig{file=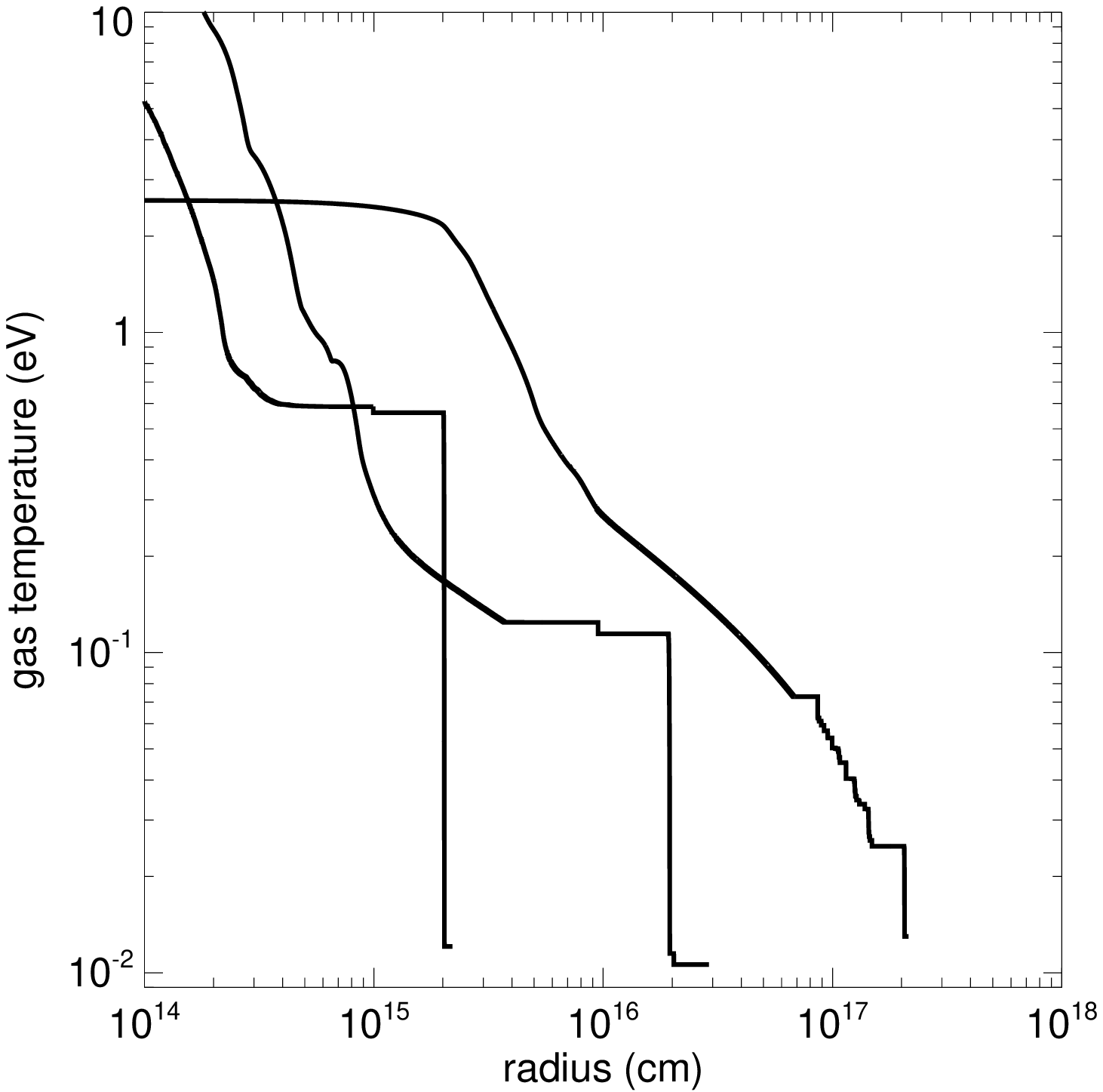,width=0.4\linewidth,clip=}  \\
\epsfig{file=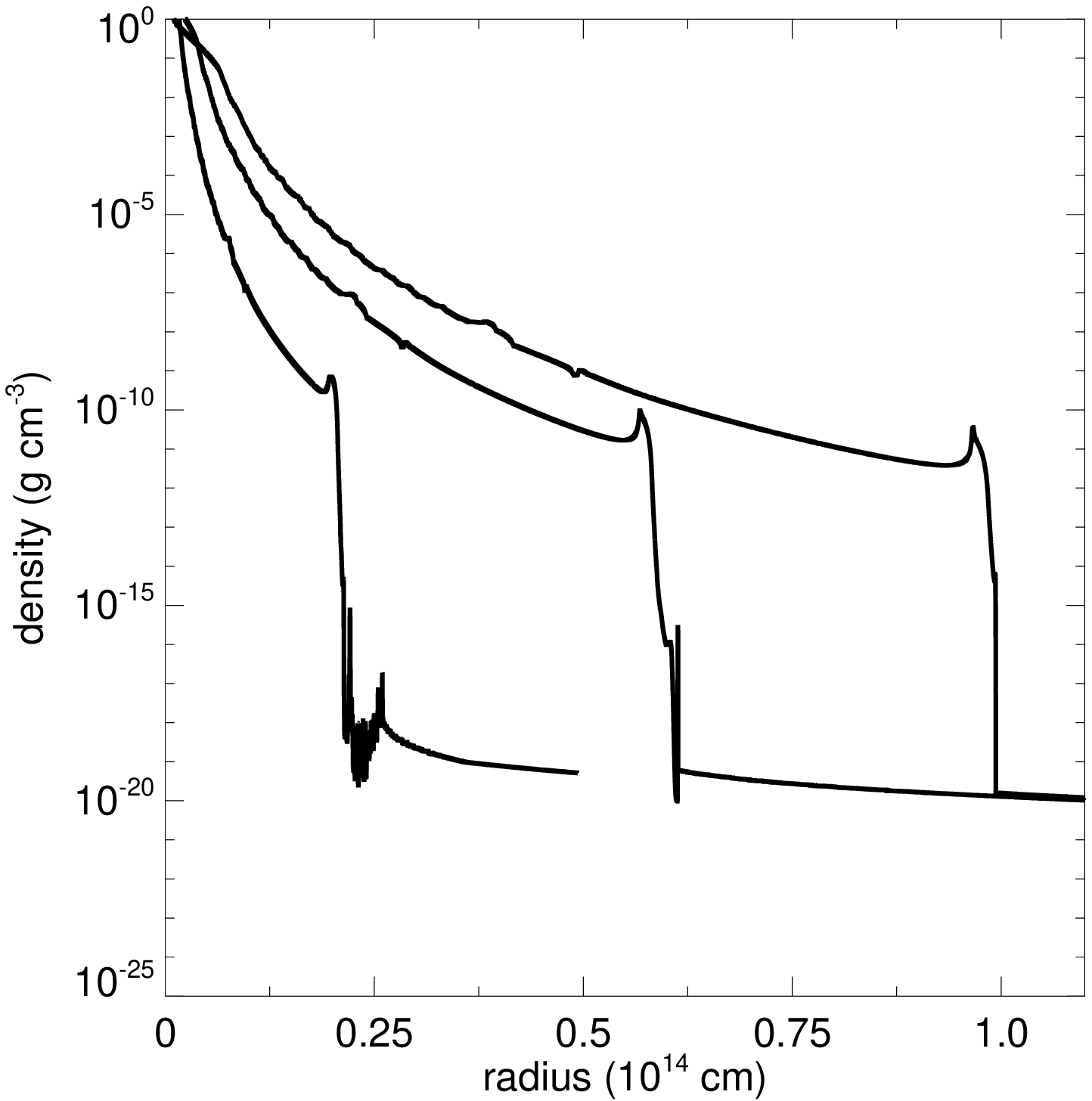,width=0.4\linewidth,clip=}  &
\epsfig{file=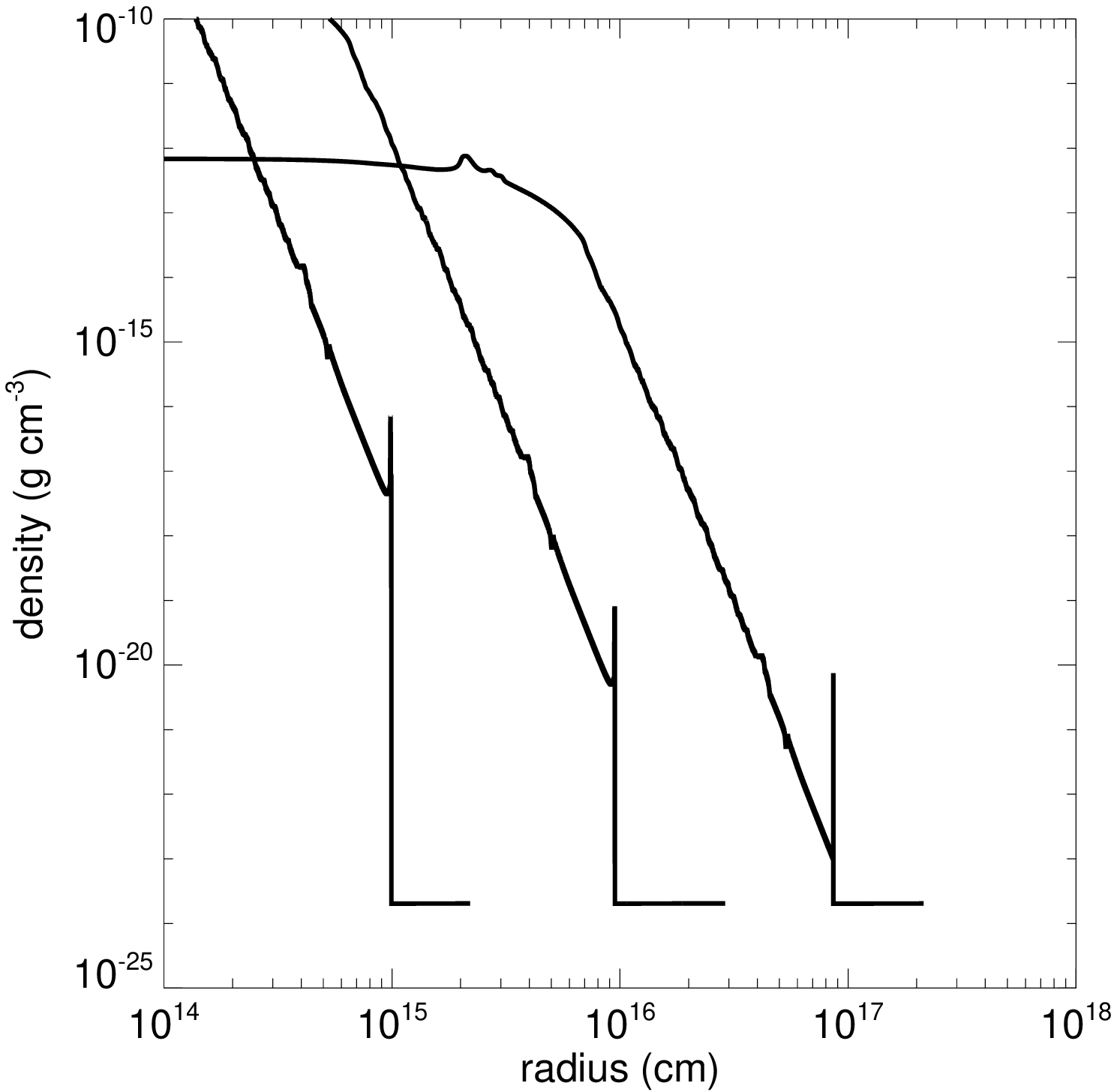,width=0.4\linewidth,clip=}
\end{tabular}
\end{center}
\caption{Early (left) and late (right) evolution of the h150 PI SN in a diffuse wind.  Left:  
from left to right the times are 121 s, 364 s and 631 s.  Right:  from left to right the times 
are 6.71 $\times$ 10$^4$ s, 6.79 $\times$ 10$^5$ s and 6.88 $\times$ 10$^6$ s.}
\label{fig:int2}
\end{figure*}

\subsection{Early and Late Evolution}

We show the early and later evolution of the h150 explosion in the left and right panels 
of Fig.~\ref{fig:int2}.  Because of the lower ambient densities, radiation from the shock 
can sustain the precursor for somewhat longer times, out to 631 s as shown in the 
density and velocity profiles.  After the collapse of the precursor, densities in the shock
fluctuate and cause the ripples in the bolometric luminosities out to $\sim$ 1000 s in 
Fig.~\ref{fig:bLC2}.  The fluctuations here are much stronger because of the lower 
densities ahead of the shock.  They are absent in the h500s0 light curve because it has 
a much lower explosion energy.  As we discuss in the next section, most of the star 
remains gravitationally bound in this SN, and only its outer layers are blown off.  There 
is not enough flux from the shock to drive fluctuations in the precursor.

After a few days the expansion of the flow is again mostly self-similar, as shown in the 
right panels in Fig.~\ref{fig:int2}.  The SN becomes a free expansion sooner in diffuse 
envelopes.  All four bolometric light curves exhibit the same rebrightening at about a 
week as in the previous light curves, and for the same reason:  the photosphere 
recedes to deeper, denser and warmer regions of the flow whose structure remains 
unaffected by the envelope.  The h150 and h200 PI SNe are brighter at late times 
because they produce much more \Ni\ than the two h500 explosions (and because the 
\Ni\ created by h500s0 again falls back into a black hole, as we discuss below).

\subsection{Light Curves}

\begin{figure}
\plotone{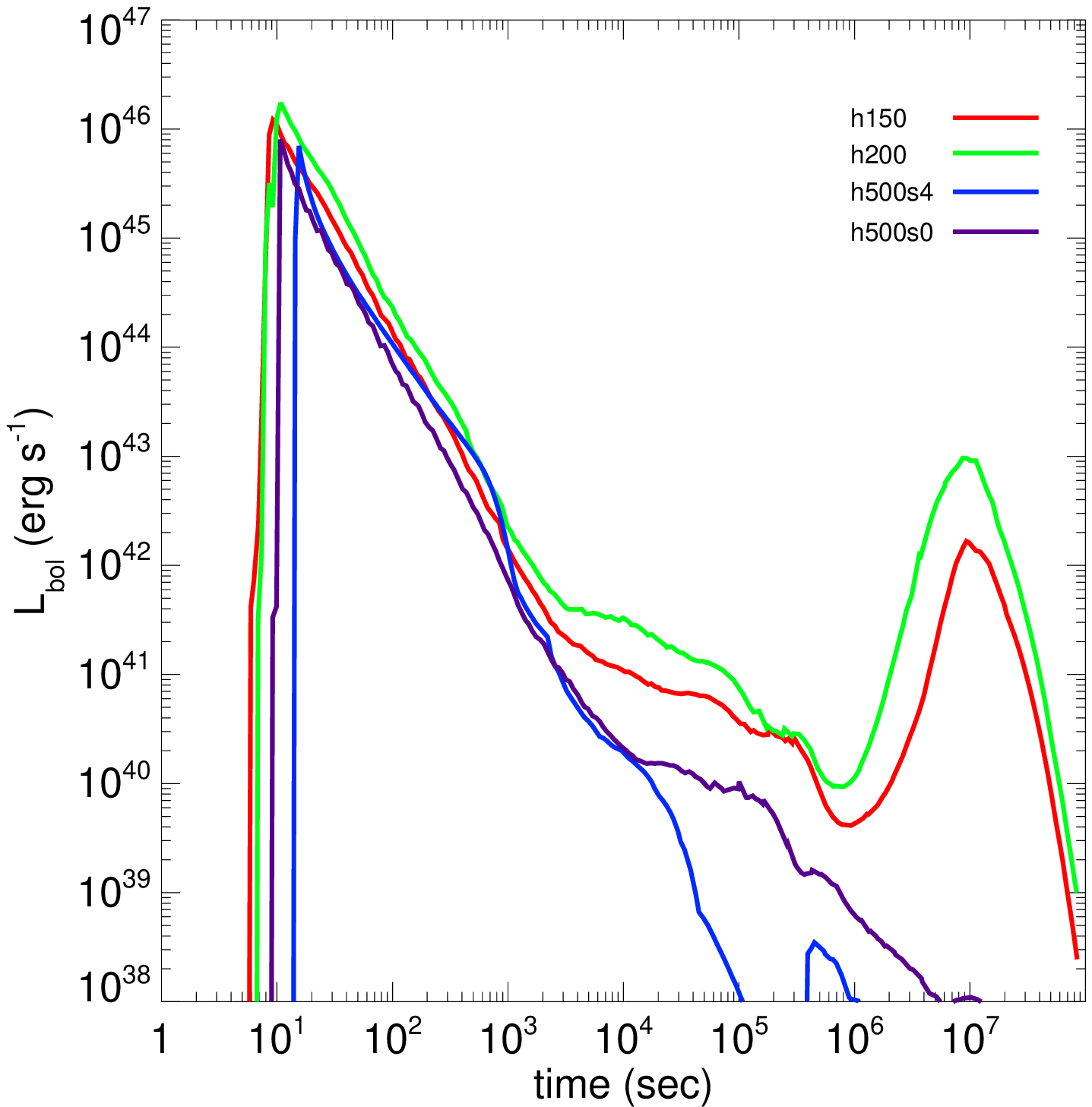}
\caption{Bolometric luminosities for all 4 explosions in diffuse envelopes.} \vspace{0.1in}
\label{fig:bLC2}
\end{figure}

Because the shock breaks out into much lower densities, peak bolometric luminosities 
range from 7 $\times$ 10$^{45}$ to 2 $\times$ 10$^{46}$ erg s$^{-1}$ for these SNe
as shown in Fig.~\ref{fig:bLC2}. They are about four orders of magnitude brighter than 
those in dense envelopes.  \Ni\ rebrightening is again evident in the h150 and h200 PI 
SNe but not in the h500 explosions for the reasons discussed above.  It reaches the 
same peak luminosity in dense and diffuse envelopes, and at about the same times.  
The presence of a dense wind therefore has little effect on \Ni\ rebrightening at later 
times because the swept up material is at too low a density to impede the escape of 
photons.  

\begin{figure*}
\epsscale{2.3}
\plottwo{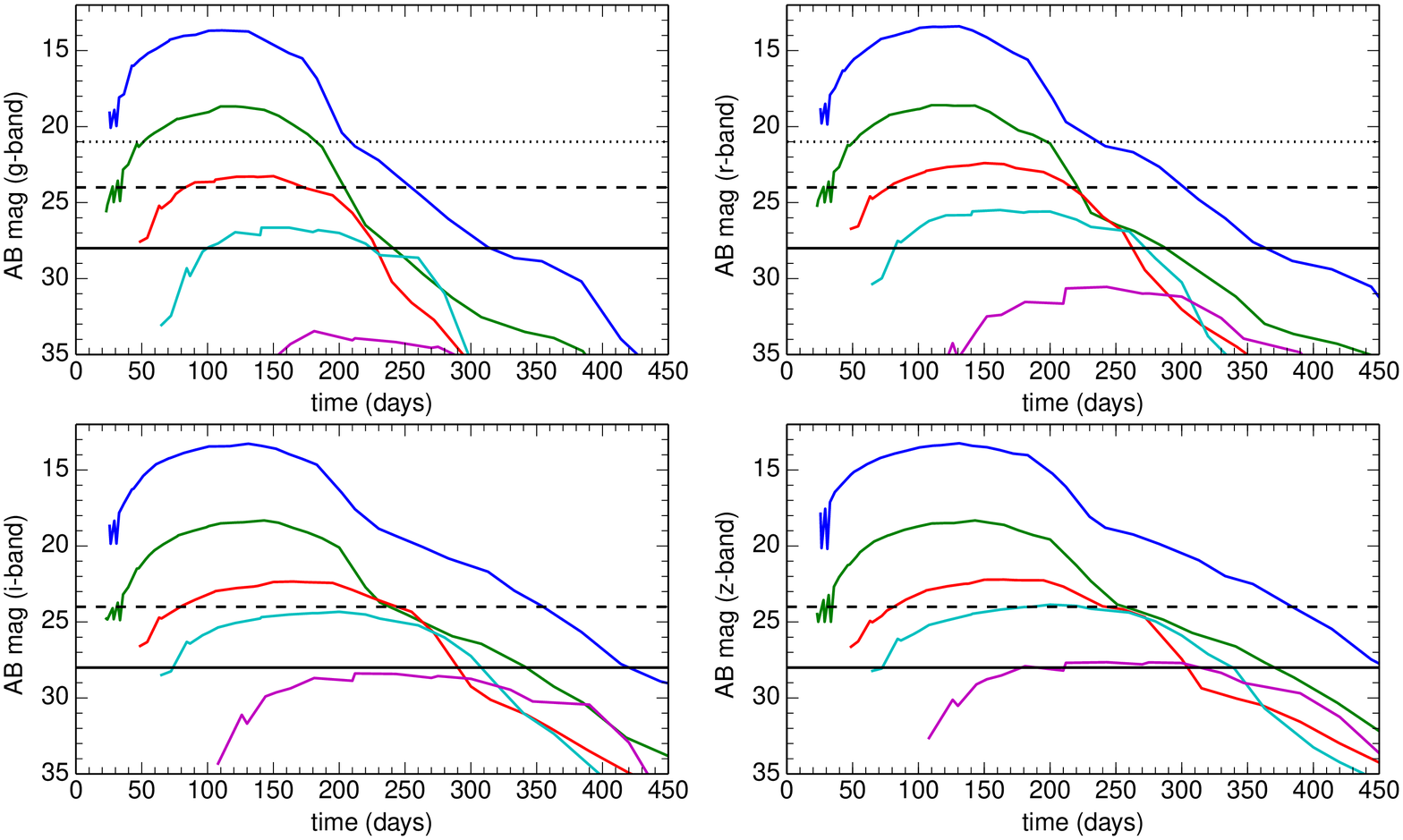}{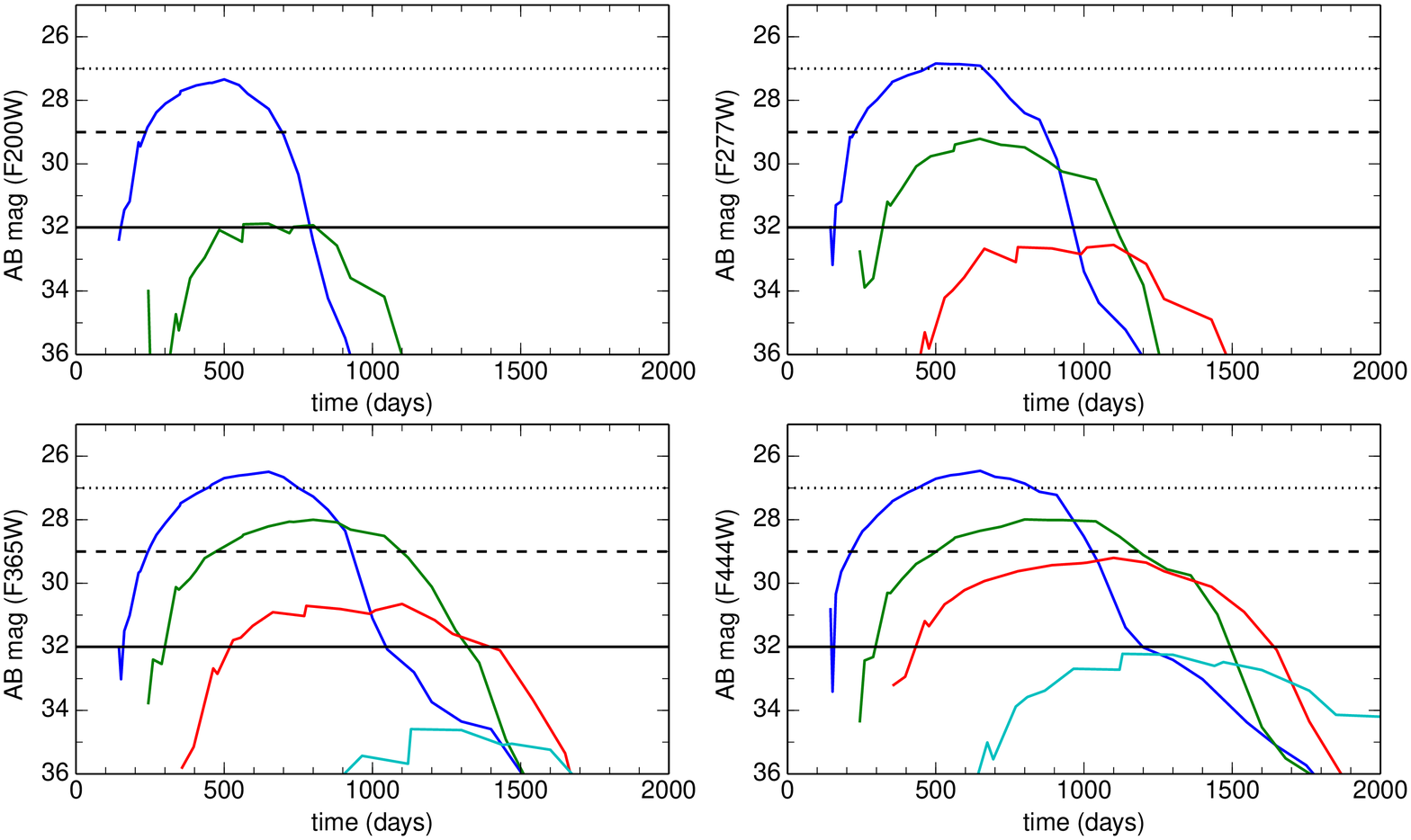}
\caption{Optical and NIR light curves for the h200 PI SN in a low-density envelope.  
Upper 4 panels:  $z =$ 0.01 (blue), $z =$ 0.1 (green), $z =$ 0.5 (red), $z =$ 1 (teal), 
and $z =$ 2 (purple).  Dotted line:  PTF detection limit; dashed line:  Pan-STARRS 
detection limit; solid line:  LSST detection limit.  Lower 4 panels:  $z =$ 4 ({\it dark 
blue}), 7 ({\it green}), 10 ({\it red}), 15 ({\it teal}), 20 ({\it purple}) and 30 ({\it yellow}).  
The horizontal dotted, dashed and solid lines are photometry limits for WFIRST, 
WFIRST with spectrum stacking and {\it JWST}, respectively.  The wavelength of 
each filter can be read from its name; for example, the F277W filter is centered at 
2.77 $\mu$m, and so forth.} \vspace{0.1in}
\label{fig:LC3}
\end{figure*}

\begin{figure*}
\epsscale{1.1}
\plotone{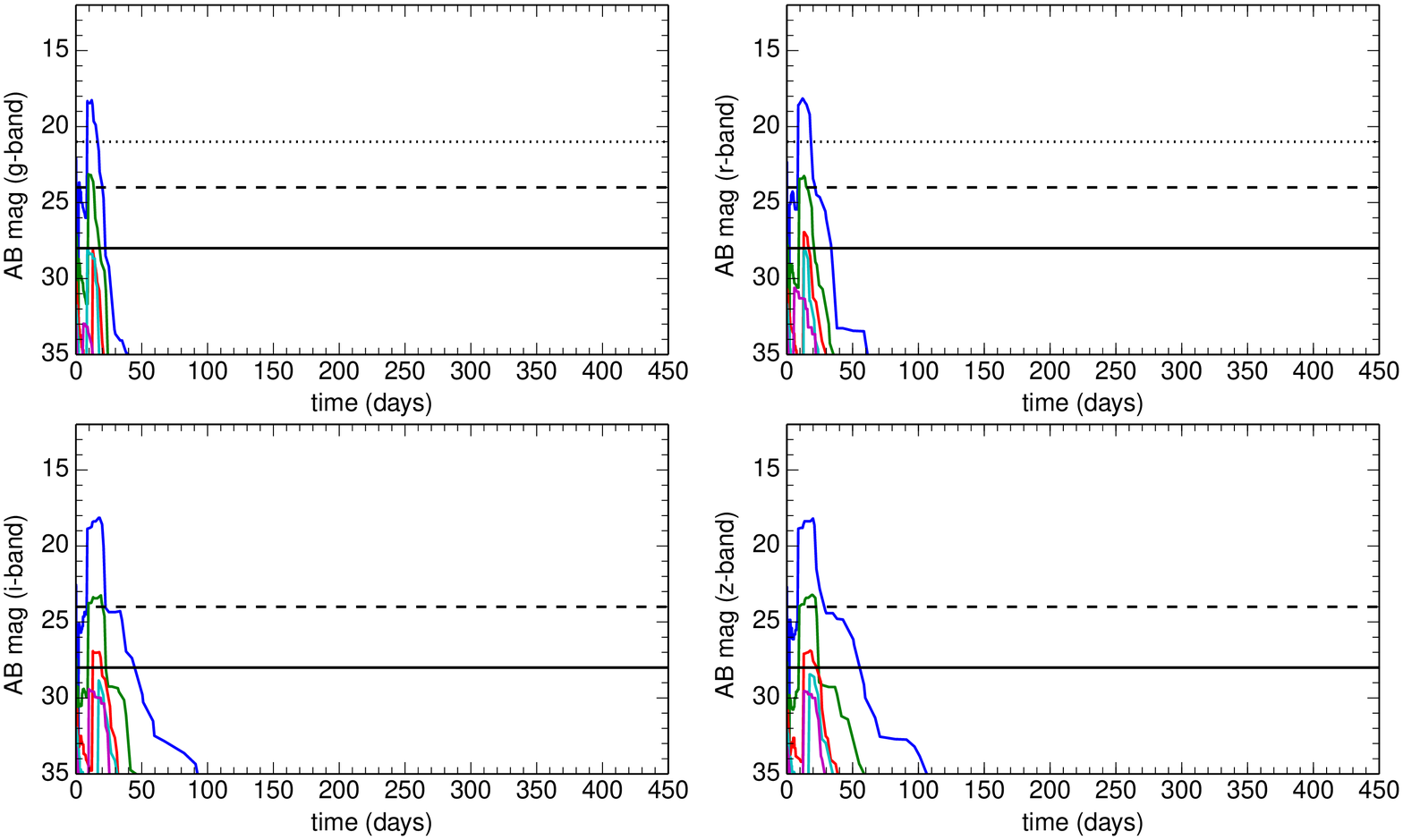}
\caption{Optical light curves for the h500s0 SN in a low-density envelope:  $z =$ 
0.01 (blue), $z =$ 0.1 (green), $z =$ 0.5 (red), $z =$ 1 (teal), and $z =$ 2 (purple).  
Dotted line:  PTF detection limit; dashed line:  Pan-STARRS detection limit; solid 
line:  LSST detection limit.} \vspace{0.1in}
\label{fig:LC4}
\end{figure*}

We show g, r, i and z band light curves for the h200 and h500s0 explosions in diffuse
envelopes in Figures~\ref{fig:LC3} and \ref{fig:LC4}.  The h200 light curves again all 
have a initial, short-lived peak due to the early expansion and cooling of the ejecta, 
which we do not show in the plots.  We instead focus on the much brighter and longer 
lived rebrightening phase, which reaches peak magnitudes that are nearly identical to 
those in the dense winds discussed in the previous section. Thus, although the dense
shroud quenches bolometric luminosities at early times it has no effect on them at 
later times or on their duration.  We again find that the h200 PI SN will be visible to 
PTF out to $z \sim$ 0.1, to Pan-STARRS out to $z \sim$ 0.5 - 1, and to LSST out to $z 
\sim$ 2.  The absence of the dense wind conversely does not enhance the brightness 
of the h500s0 explosion; it is still a much dimmer event, again only being visible out to 
$z \sim$ 0.01 to PTF, $z \sim$ 0.1 to Pan-STARRS and $z \sim$ 0.5 to LSST.

We find that whether the envelope is dense or diffuse has little impact on detection 
limits in the NIR for the h200 explosion.  NIR light curves at 2 - 4 $\mu$m for h200 at 
$z =$ 4, 7, 10, and 15 are shown in Fig.~\ref{fig:LC3}.  We again find that this SN will 
be visible to WFIRST out to $z \sim$ 4 - 10 and to {\it JWST} out to $z \sim$ 10 - 15, 
the era of first galaxy formation.  The low-density wind likewise does not improve 
detection limits for the more massive h500s0 and h500s4 PI SNe in the NIR, they are 
still well below even the sensitivity of {\it JWST} at $z >$ 4.

\section{Fallback and Black Hole Production}

\begin{figure*}
\plottwo{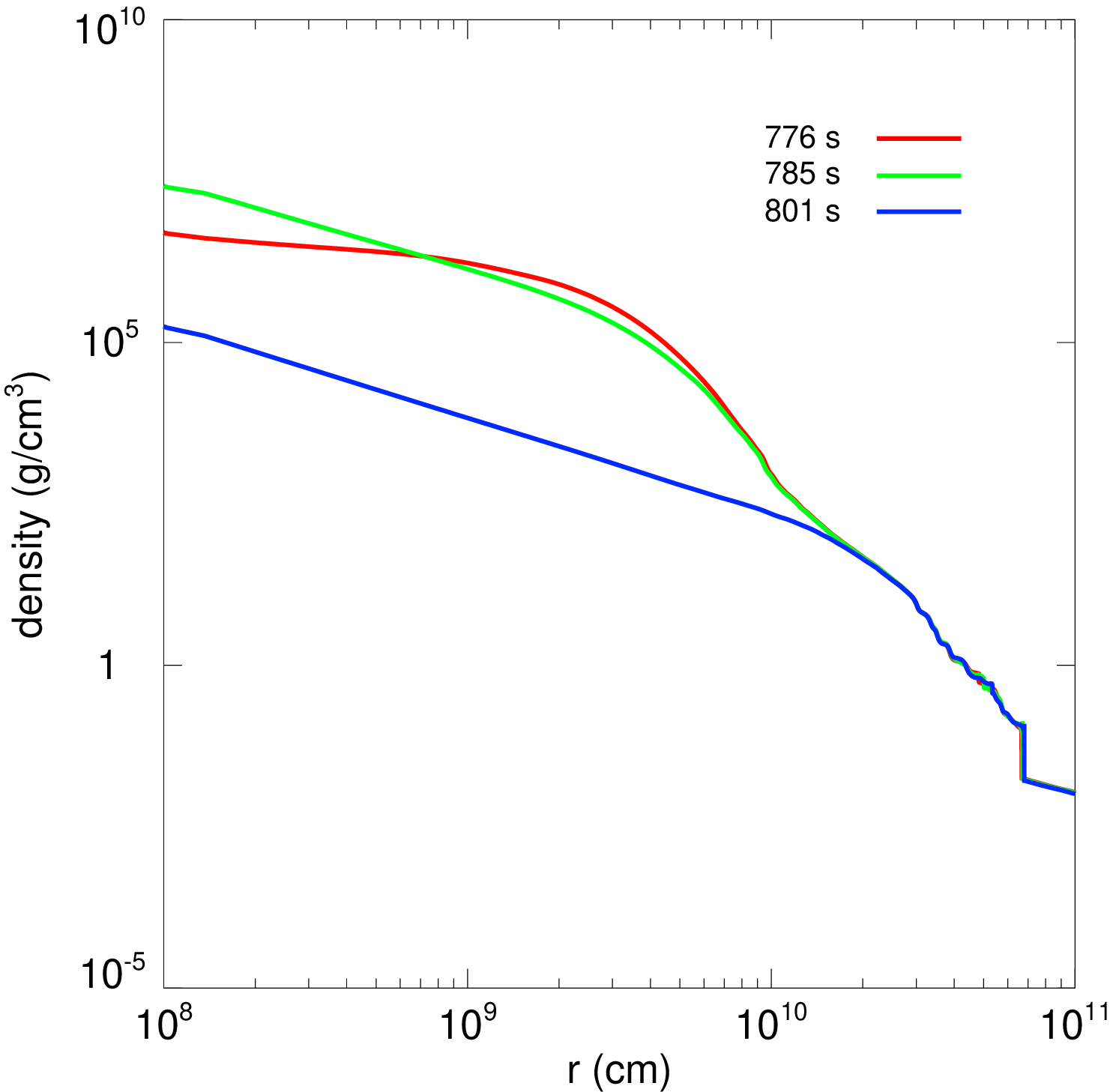}{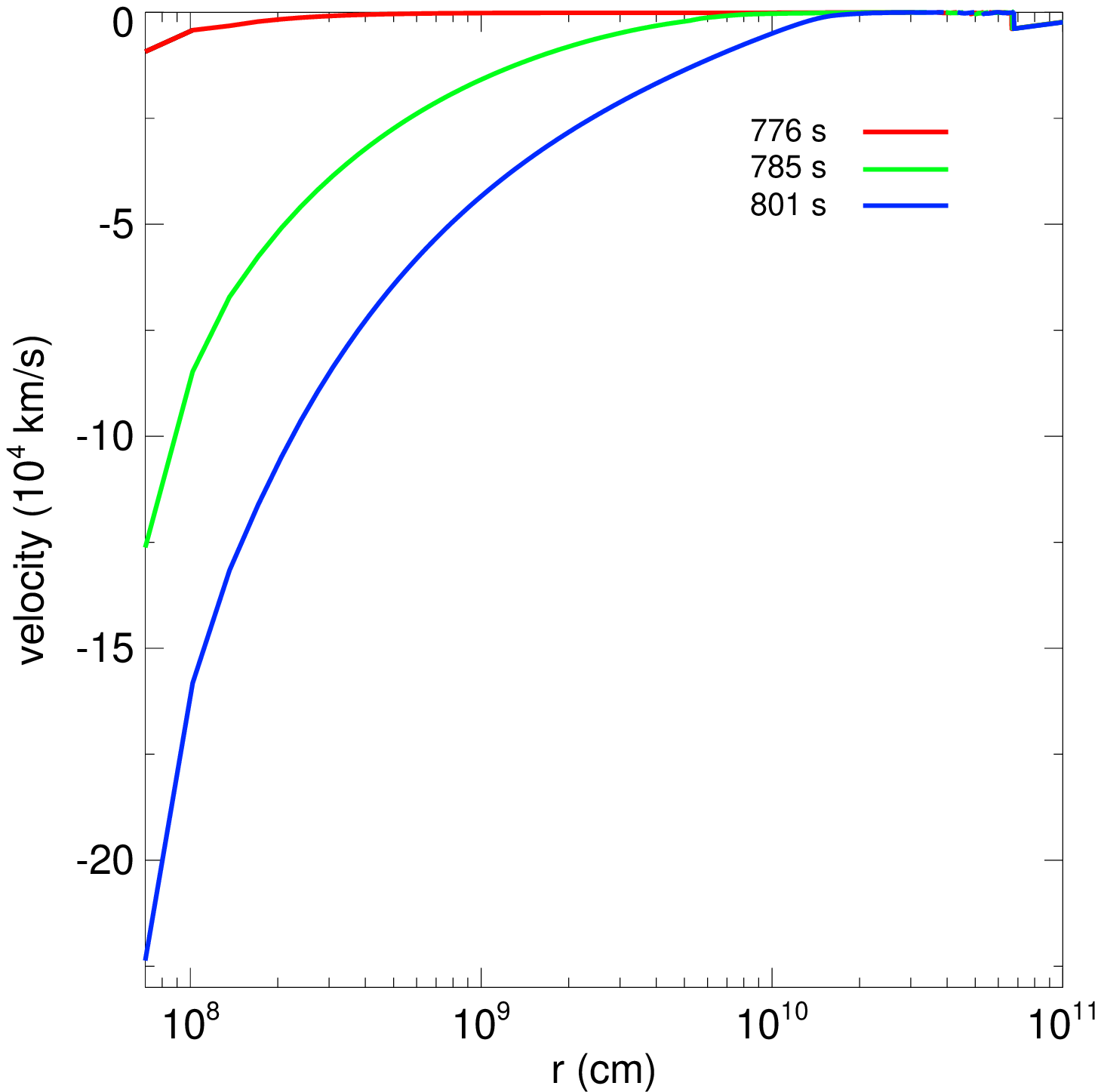}
\caption{Fallback in the h500s0 explosion.  Left:  densities.  Right: velocities.} 
\label{fig:fb}
\end{figure*}

\begin{figure}
\plotone{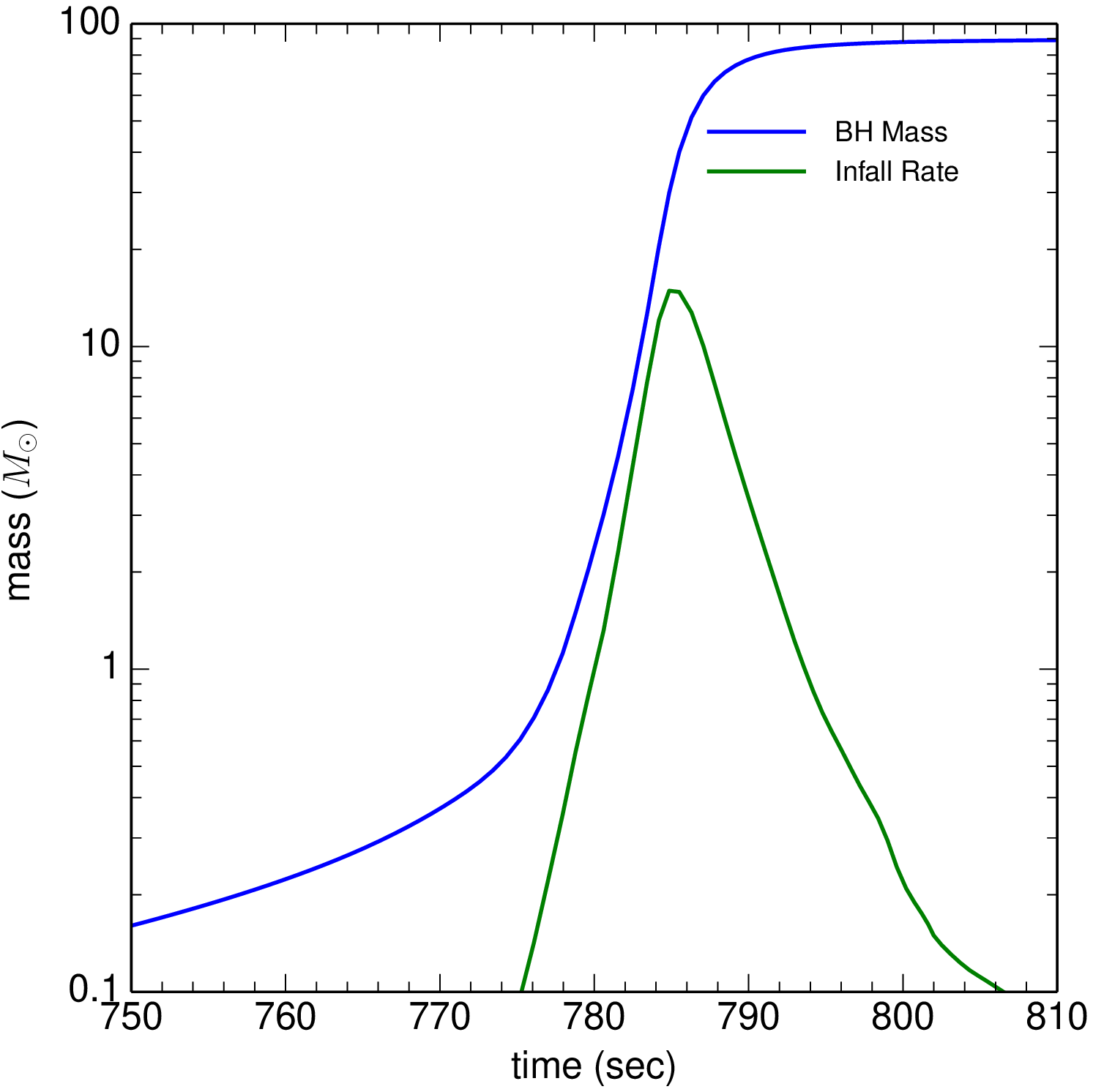}
\caption{Central black hole mass and infall rates in the h500s0 PI SN.} \vspace{0.1in}
\label{fig:bh}
\end{figure}

PI SNe are usually thought to completely unbind the star and leave no compact remnant.
Here, we report that some PI SNe do create massive black holes.  In Fig.~\ref{fig:fb} we 
show density and velocity profiles at the center of the h500s0 explosion, which is the 
weakest of the set.  Because this SN has a net energy of only 3.7 $\times$ 10$^{51}$ erg,
most of the material interior to 10$^{11}$ cm remains gravitationally bound.  At 776 s it 
has come to a halt, and the deepest regions begin to fall back to the center of the grid. By 
785 s $\sim$ 30 \Ms\ has fallen back to the center at peak velocities of 12,000 km s$^{-1}
$ and infall rates of $\sim$ 15 \Ms\ s$^{-1}$, as shown in Fig.~\ref{fig:bh}. By 801 s, $\sim
$ 90 \Ms\ has collapsed to the center of the grid.  In our model we neglect radiative 
feedback from the nascent neutron star and black hole, which could regulate fallback rates 
and contribute to the luminosity of the explosion, so these infall rates should be taken to be 
upper limits.

The \Ni\ formed during the explosion quickly collapses to form a neutron star as 
infalling material from above crushes down onto it.Ê Could this weak PI SN create the 
conditions for a post outburst core-bounce SN?Ê The fate of the infalling mass depends 
upon the accretion rate onto the newly formed neutron star \citep{fryer99,fry12}.Ê The 
accretion rate determines the ram pressure that the SN engine must overcome to drive 
an explosion.  If it is too high, the convective engine is unable to push off the accreting 
material, and the neutron star gains mass until it collapses to form a black hole.Ê For 
typical 15 \Ms\ stars, the accretion rate during the engine phase drops below 1 \Ms\ 
s$^{-1}$ in the first 100 ms.Ê When it drops below this value, strong SN explosions are 
expected \citep{fryer99}.Ê In more massive stars, the accretion rate never falls below 1 
\Ms\ s$^{-1}$ (in a 25 \Ms\ star it drops to just above 1 \Ms\ s$^{-1}$ in the first second 
and to 2.5 \Ms\ s$^{-1}$ in a 40 \Ms\ star).  These systems are usually thought to have 
weak or no SN explosions \citep{fryer99,fry12}.Ê In our model, the accretion rate peaks 
at $\sim$ 10 \Ms\ s$^{-1}$, so the neutron star will almost certainly collapse to a black 
hole without an explosion.Ê 

However, if the infalling material has enough angular momentum, this system has 
accretion rates that are high enough to power a standard collapsar gamma-ray burst 
\citep[GRB;][]{pop99}.Ê If a GRB is produced, a secondary explosion with a distinctive 
gamma-ray signature might accompany the PI SN.  We do not expect the h500s0 model
to produce a GRB because its progenitor has no rotation.  By 801 s, fallback is complete,
and a 90 \Ms\ black hole has formed at the center of the grid.

\section{Conclusion}

It is clear that there is far more variety to PI SNe and their light curves than previously 
thought, and that Pop III PI SNe can no longer be considered to represent the population 
as a whole, particularly at higher metallicities.  We find that non-zero metallicity PI SNe 
in general are dimmer than Pop III events and are only visible at lower redshifts.  But 
some of them, like the h150 and h200 SNe, will be bright enough to be detected at the 
earliest epochs at which they can occur, the formation of the first galaxies, and can be 
used to probe the properties of stars in those galaxies.  Others, like the h500 explosions,
are only visible in the local universe and could be mistaken for a variety of dim CC SNe.
Metals lead to drastic departures from Pop III stellar evolution and final structures for the 
star, and they also give rise to complex circumstellar media. Because the structure of the 
ejecta and its surroundings can both radically alter the light curve of the explosion, PI 
SNe at near-solar metallicities cannot be approximated by simply stripping the H envelope 
from a Pop III star and then exploding it, as has been done in previous studies.  The 
effects of metallicity on the evolution of the star from birth and its ambient medium at later 
times must be taken into account to properly identify these events in the local universe.  

While our prescription for mass loss from our massive PI SN progenitors is a reasonable
one, it should not be considered to be comprehensive.  There could be violent episodes 
of mass loss prior to explosion due to vigorous convection or other mechanisms 
\citep[e.g., gravity waves; see][]{qs12}.  Common envelope evolution in binaries is often
observed with massive stars and can lead to highly anisotropic mass ejections \citep{se11,
chini12}. Besides radically altering the environment of the blast, and hence its light curve,
these processes can also change the evolution of the star itself.  Furthermore, although
we do consider explosions in dense envelopes, these envelopes do not contain all the
mass lost by the star over its life, and if this loss was isotropic it would present much 
higher optical depths to the SNe than those considered here.  However, the mass loss is 
likely anisotropic \citep[i.e.,][]{pas12}, and large solid angles in the sky could have fairly 
low optical depths to the SN shock.  Our luminosities should be considered to be upper 
limits to the those for such events.

We emphasize that because only a small subset of the PI SNe that are possible 
have been studied to date, it is premature to dismiss them as candidates for a few 
recent superluminous SNe (SLSNe) just because their light curves and spectra do not 
match those of Pop III explosions \citep[e.g.,][]{det12,nic13} \citep[see also][for an 
alternate interpretation of SN 2007bi]{det12a}.  For example, we have not examined 
$Z \sim$ 0.1 \Zs\ PI SNe crashing into dense shells ejected by the star prior to its death.  
Similar but less energetic Type IIn events have been found to have rapid rise times 
consistent with SLSNe \citep{wet12e}. Had the progenitors in this study ejected massive 
shells at later times that were within reach of their ejecta, they might have had light 
curves that are consistent with the superluminous events in \citet{nic13}, and they may 
have been observable at much higher redshifts than the explosions in this paper.  Such 
collisions are now being studied with RAGE. The large number of possibilities for PI SN 
progenitor structure and envelope also highlights the difficulty of matching any one PI 
SN candidate to current models \citep[see also][]{bay14}.  That said, we note that the 
bolometric light curve for SN 2007bi has now been well reproduced by the PI SN of a 
low-metallicity 250 \Ms\ star, although it remains to be seen if its spectra are also a 
match \citep{kz14a}.

Our weak pair-instability supernovae, in which much of the \Ni\ falls back, produce dim, 
short-duration outbursts.Ê A growing list of dim supernovae have been observed, often 
being classified as thermonuclear flashes on white dwarfs, the so-called ".Ia supernovae" 
because they are 1/10th the strength of normal supernovae \citep[e.g.,][]{drout13}.Ê
Many of these events have been explained as CC SNe \citep{fet07,fet09,moriya10b,kk13}.Ê 
But our models show that some PI SNe could also be masquerading as dim supernovae.Ê 
If PI SNe can explain both superluminous and dim supernovae, it is possible that, under 
the right conditions (explosion energy and circumstellar medium), they could be hidden 
in a wide variety of supernova classes.

Rotation can have the same effect as metallicity on PI SN light curves. \citet{yoon12}, \citet{
cw12}, and \citet{cwc13} find that Pop III stars that rotate at 50\% of the breakup velocity at 
their equator can lose mass and explode as PI SNe at 85 - 135 \Ms.  They, like the stars in 
this paper, shed their H envelopes and die as bare He cores with explosion energies and 
breakout radii that are similar to those here. They also can explode in dense winds or shells.
We find that these too can range from being rather dim events to being visible in the era of
first galaxy formation \citep{aj06,smidt14a}.  

Two of our stars have metallicities slightly above the maximum predicted for PI SN 
progenitors by some studies \citep[0.33 \Zs;][]{lang07} and, for now, should be considered 
to be upper limits for the metallicities of these events.  There is a degeneracy between 
progenitor mass, metallicity, mass loss rate and rotation in light curves for these PI SNe 
that makes it difficult to predict their frequency.  For example, a star with an initial mass 
below 300 \Ms\ and a lower mass loss rate could end up with a final mass and internal 
structure similar to that of one of our 500 \Ms\ models and explode with nearly the same 
energy.  Given these degeneracies, this type of explosion might be more common than 
would be suggested by the number of 400 - 500 \Ms\ stars alone.

Our models demonstrate, for the first time, that some PI SNe do create massive black holes,
like the 90 \Ms\ remnant created in the h500s0 explosion.  They also reveal that fallback can
strongly affect the luminosity of the SN because any \Ni\ the explosion creates is swallowed 
up by the black hole at early times and therefore cannot power the light curve at later times.
The creation of a black hole also admits the possibility of the formation of a black hole 
accretion disk system, if the core of the star has sufficient angular momentum.  Because in
excess of 1 \Ms\ can fall into the black hole every second, a small fraction of these events 
could be accompanied by GRBs \citep{wet08c,nak12,mes13a} because the star has shed 
its H envelope and the jet can escape.

The SN factories may ultimately discover many PI SNe, even though they can only detect 
them at $z \lesssim$ 2, because cosmic star formation rates are so much higher at low 
redshifts than at high redshifts \citep{tfs07,ts09,idf11,camp11,re12,cooke12,wise12,jdk12,
pmb12,xu13,haseg13,mura13}.  Indeed, \citet{lang07} estimate that there is one non-zero 
metallicity PI SN per thousand SNe today and as many as one per one hundred at $z =$ 5
\citep[see also][]{lang09}.  The new SN facilities, together with future NIR observatories, will 
soon probe the stellar populations of both early and mature galaxies with these energetic 
transients.

\acknowledgments

D.J.W. was supported by the European Research Council under the European Community's 
Seventh Framework Programme (FP7/2007-2013) via the ERC Advanced Grant "STARLIGHT: 
Formation of the First Stars" (project number 339177). Both he and C.J. are grateful for support 
by the DOE Institute for Nuclear Theory during the Extreme Computing Workshop INT-11-2a, 
where some of this work was performed. He also thanks Terrance Strother for running some of 
the models.  R.H. was supported by the World Premier International Research Center Initiative 
(WPI Initiative), MEXT, Japan and acknowledges support from EU-FP7-ERC-2012-St Grant 
306901.  N.Y. acknowledges support from the Ministry of Higher Education and University of 
Malaya under the Higher Education Academic Training Scheme and the Commonwealth 
Scholarship Commission for the Split-Site PhD 2010 - 2011 programme at the University of Keele 
and acknowledges support from Ministry of Education of Malaysia under FRGS FP003-2013A.  
A.H. and K.C. were funded by the US Department of Energy under contracts DE-FC02-01ER41176, 
FC02-09ER41618 (SciDAC), and DE-FG02-87ER40328, and A.H. also acknowledges support by a 
Future Fellowship (FT120100363) from the Australian Research Council.  Work at LANL was done 
under the auspices of the National Nuclear Security Administration of the U.S. Dept of Energy at 
Los Alamos National Laboratory under Contract No. DE-AC52-06NA25396.  All RAGE and 
SPECTRUM calculations were performed on Institutional Computing (IC) platforms at LANL 
(Mustang, Pinto and Lobo).  

\bibliographystyle{apj}
\bibliography{refs}

\end{document}